\documentclass[aps,prb,twocolumn,superscriptaddress,nofootinbib,showpacs]{revtex4}
\usepackage{amssymb}
\usepackage{graphicx}
\usepackage{epsf,epsfig}
\usepackage{bm}
\usepackage{bbm}
\usepackage{amsfonts}
\usepackage{amsmath}
\usepackage{graphics}
\usepackage[normalem]{ulem}
\usepackage{color}

\newcommand{\be}{\begin{eqnarray}}
\newcommand{\ee}{\end{eqnarray}}
\newcommand{\ba}{\begin{array}}
\newcommand{\ea}{\end{array}}
\newcommand{\no}{\nonumber}
\newcommand{\tr}{\mbox{tr}}
\newcommand{\Tr}{\mbox{Tr}}

\newcommand{\eps}{\varepsilon}

\newcommand{\bfr}{{\bf r}}

\newcommand{\bfq}{{\bf q}}
\newcommand{\bfk}{{\bf k}}

\newcommand{\bfj}{{\bf j}}

\begin{document}

\title{Heat diffusion in the disordered electron gas}

\author{G. Schwiete}
\email{schwiete@uni-mainz.de} \affiliation{Spin Phenomena Interdisciplinary Center (SPICE) and Institut f\"ur Physik,
Johannes Gutenberg Universit\"at Mainz, 55128 Mainz, Germany}
\affiliation{Dahlem Center for Complex Quantum Systems and Institut f\"ur Theoretische
Physik, Freie Universit\"at Berlin, 14195 Berlin, Germany
}
\author{A. M. Finkel'stein}
\affiliation{Department of Physics and Astronomy, Texas A\&M University, College Station, TX 77843-4242, USA}
\affiliation{Department of
Condensed Matter Physics, The Weizmann Institute of Science, 76100
Rehovot, Israel}
\affiliation{L. D. Landau Institute for Theoretical Physics, 117940 Moscow, Russia}

\begin{abstract}
We study the thermal conductivity of the disordered two-dimensional electron gas. To this end we analyze the heat density-heat density correlation function concentrating on the scattering processes induced by the Coulomb interaction in the sub-temperature energy range. These scattering processes are at the origin of logarithmic corrections violating the Wiedemann-Franz law. Special care is devoted to the definition of the heat density in the presence of the long-range Coulomb interaction. To clarify the structure of the correlation function, we present details of a perturbative calculation. While the conservation of energy strongly constrains the general form of the heat density-heat density correlation function, the balance of various terms turns out to be rather different from that for the correlation functions of other conserved quantities such as the density-density or spin density-spin density correlation function.
\end{abstract}

  \maketitle

\section{Introduction}
While experimentally thermal transport is controlled by boundary conditions, for the theoretical description it is more convenient to study the response to a gradient of temperature. A principle difficulty for the description of thermal transport is that a temperature gradient does not correspond to an external "mechanical" force like the one originating from an electric potential. To bypass this problem, time-dependent "gravitational potentials" can be introduced \cite{Luttinger64,Shastry09,Michaeli09,Schwiete14a,Schwiete14b} as source fields in the microscopic action. The heat density-heat density correlation function can be found by a variation of the action with respect to these source fields. Knowledge of the correlation function allows to determine the thermal conductivity.\cite{Castellani87,Schwiete14a,Schwiete14b}

An unpleasant difference of the gravitational potentials with respect to, for example, electromagnetic potentials, is that the gravitational potentials couple to all terms constituting the Hamiltonian density. This includes, in particular, the interaction part. Furthermore, in the presence of impurities, the gravitational potentials also couple to the disorder part of the Hamiltonian. In Ref.~\onlinecite{Castellani87}, the latter problem has been overcome by a special diagrammatic procedure. Recently, we showed how the use of the gravitational potentials can be merged with the NL$\sigma$M formalism, and performed a renormalization group (RG) analysis for the thermal conductivity of a disordered Fermi liquid system with short-range interaction potentials.\cite{Schwiete14a,Schwiete14b} The RG procedure covers the interval of energies with the elastic scattering rate $1/\tau$ as the upper cutoff and the temperature $T$ as the lower one ($T\ll 1/\tau$).

Combined measurements of thermal and electric conductivites are often employed in order to assess the applicability of the quasiparticle description.\cite{Tanatar07,Smith08,Pfau12,Pfau13,Mahajan13,Dong13,Sutherland15,Taupin15} The analysis of Refs.~\onlinecite{Schwiete14a,Schwiete14b} revealed that for the two-dimensional disordered system with short range interactions the Wiedemann-Franz law\cite{Wiedemann1853} (WFL) holds even in the presence of quantum corrections caused by the interplay of diffusion modes and the electron-electron interaction. Generally speaking, the WFL should not be considered as a strict law outside the realm of single-particle physics. This is already evident from the very fact that the potential used for calculating the electric conductivity couples to the particle density only, while the gravitational potential probes the entire Hamiltonian density. Still, the RG analysis shows that at least for the leading logarithmic corrections in a two-dimensional system with short range interactions, the WFL is obeyed.

In this paper, we present a perturbative analysis of logarithmic corrections to the heat-density heat-density correlation function in a two-dimensional electron gas, i.e., in a system with long-range Coulomb interaction. Since the effects of the Coulomb interaction in the RG-interval of energies have already been studied in perturbation theory\cite{Castellani87} and are very similar to the case of the short-range interaction,\cite{Schwiete14a,Schwiete14b} we will focus our attention on the sub-temperature energy range, which is beyond the scope of the RG analysis. The main difference between the RG-interval and sub-temperature energy range is that, while the transitions described by the standard RG procedure are virtual, the sub-thermal range deals with the \emph{on-shell} scattering. For the analysis of the logarithmic corrections to electric conductivity, the sub-thermal processes can usually be neglected. Thermal conductivity constitutes an important exception. Here, the scattering processes induced by the long-range Coulomb interaction yield logarithmic corrections which, in principle, may compete with the RG-corrections. The corrections caused by the \emph{on-shell} scattering, in contrast to those of the RG origin, violate the WFL. In this manuscript, we identify the relevant diagrams and find the correction to the WFL. We show how the terms violating the WFL become compatible with the general form of the heat density-heat density correlation function. We thereby demonstrate the consistency of our results with the general scheme for the calculation of a correlation function of the density of a conserved quantity, which in our case is the energy.

Our study differs from previous related work\cite{Castellani87,Livanov91,Arfi92,Raimondi04,Niven05,Catelani05,Catelani07,Michaeli09} in several respects. The heat-density heat-density correlation function was studied before in Ref.~\onlinecite{Castellani87}. However, logarithmic corrections originating from the sub-temperature regime were not taken into account in this work. Other studies of thermal conductivity available in the literature can be divided into Kubo-type linear response calculations based on the heat-current heat-current correlation function\cite{Arfi92,Niven05} and kinetic equation approaches.\cite{Livanov91,Catelani05,Catelani07,Michaeli09} Our final result for the thermal conductivity of the system with Coulomb interaction coincides with the one stated in Refs.~\onlinecite{Raimondi04,Niven05,Catelani05,Michaeli09}. While the mentioned works arrived at the same final result, they did not agree on the definition of the heat density and of the associated heat current, a question of principle importance for the calculation of the thermal conductivity. We will devote special attention to this point.

The paper is organized as follows. In Sec.~\ref{sec:model} we state general properties of the heat-density heat-density correlation function as well as its relation to the quantity of our interest, the thermal conductivity. We also introduce gravitational potentials as source fields in the action. In Sec.~\ref{sec:CoulombNLSM} we define the heat density for the electron system with Coulomb interaction and present the NL$\sigma$M in the presence of the gravitational potentials. This model will serve as a starting point for the calculation of the heat-density heat-density correlation function. The Coulomb problem has a peculiar feature: While we are interested in heat transport in a two-dimensional electron system, the natural definition of a local conservation law connecting heat density and heat current requires a three-dimensional setting. The reason is that a part of the energy of the system is stored in the electromagnetic field, and this field is not restricted to the two-dimensional plane. In order to define transport of heat in two dimensions, we devise a specific projection procedure. Special care has to be taken already on the level of the definition of the three-dimensional conservation law. The principle of gauge invariance plays a pivotal role in unambiguously identifying the heat density and heat current. In the present context, this aspect was stressed in Appendix B of Ref.~\onlinecite{Catelani05}. We illuminate this point further in Appendix~\ref{sec:belinfante}, where we stress the connection with the field theoretic construction of the Belinfante energy-momentum tensor.\cite{Belinfante40,Greiner96} In Sec.~\ref{sec:Dynamic} we collect basic formulas required for the calculation of the dynamical part of the heat-density heat-density correlation function. Further on, in Sec.~\ref{sec:gstructure} 
we discuss the structure of the static and dynamic parts of the heat-density heat-density correlation function. Special attention is devoted to the consistency with the constraint imposed by the energy conservation law. Finally, in Sec.~\ref{sec:DynamicP} we introduce the diagrammatic representation and present the general analysis of logarithmic corrections for the heat density-heat density correlation function. In particular, we clarify how corrections from the sub-temperature interval of energies (caused by on-shell scattering processes) enter the heat density correlation function and modify the thermal conductivity. The full list of logarithmic contributions of various kind is given in Appendix~\ref{sec:listlogs}. Throughout Secs.~\ref{sec:Dynamic}%\ref{sec:gstructure} and
-\ref{sec:DynamicP} as well as in Appendix~\ref{sec:listlogs}, we systematically compare the heat density-heat density correlation function with the well-studied example of the density-density correlation function in order to stress differences and similarities.

\section{Generalities: thermal conductivity and the heat-density correlation function}
\label{sec:model}

In this work, we use the Keldysh technique\cite{Schwinger61,Kadanoff62,Keldish65,Kamenev11}, which allows us to calculate the correlation function directly in real time. The action is defined on the Keldysh time-contour $\mathcal{C}$ consisting of forward ($+$) and backward ($-$) branches. We start our considerations with the action
\be
S_{k}[\psi^\dagger,\psi]=\int_\mathcal{C}dt \int_{\bfr} \left(\psi^\dagger i \partial_t\psi-k[\psi^\dagger,\psi]\right),
\ee
which contains the heat density $k$ explicitly. The heat density is defined as $k=h-\mu n$, where $h$ and $n$ are the hamiltonian density and particle density and $\mu$ is the chemical potential. Further, $\psi=(\psi_\uparrow,\psi_\downarrow)$, $\psi^\dagger=(\psi^*_\uparrow,\psi^*_{\downarrow})$ are Grassmann fields with two spin components.

We wish to calculate the retarded heat density correlation function $
\chi_{kk}(x_1,x_2)=-i\theta(t_1-t_2)\langle[\hat{k}(x_1),\hat{k}(x_2)]\rangle_T$, where $x=(\bfr,t)$, $\hat{k}=\hat{h}-\mu\hat{n}$ is the heat density operator and the angular brackets denote thermal averaging. The definition of the heat density operator reflects the fact that we study the propagation of heat under the condition when mechanical work (e.g. the radiation of acoustic waves) can be neglected. For the calculation we define the classical ($cl$) and quantum components ($q$) of the heat density symmetrized over the two branches of the Keldysh contour, $k_{cl/q}=\frac{1}{2}(k_+\pm k_-)$, \cite{Kamenev11} and write the retarded correlation function as $\chi_{kk}(x_1,x_2)=-2 i\left\langle k_{cl}(x_1) k_q(x_2)\right\rangle$, where the averaging is with respect to the action $S_k$.

In order to generate the heat density correlation functions, we add the source term
\be
S_\eta=2\int_x [\eta_2(x)k_{cl}(x)+\eta_1(x)k_q(x)].\label{eq:Seta}
\ee
to the action. Then, one can find $\chi_{kk}$ as
\be
\chi_{kk}(x_1,x_2)&=&\left.\frac{i}{2}\frac{\delta^2 \mathcal{Z}}{\delta \eta_2(x_1)\delta \eta_1(x_2)}\right|_{\eta_2=\eta_1=0},\label{eq:chisources}
\ee
where $\mathcal{Z}=\int D[\vec{\psi}^\dagger,\vec{\psi}]\exp(iS_k+iS_\eta)$ is the partition function and $\vec{\psi}$ and $\vec{\psi}^\dagger$ generalize the definition of $\psi$ and $\psi^\dagger$ to the Keldysh space. The so-called gravitational potentials $\eta_1$ and $\eta_2$ in $S_\eta$ allow us to formulate a linear response theory for the heat transport.

To find the thermal conductivity, it remains to establish a connection between the response to the gravitational potential and the response to a temperature variation $\delta T$. As argued by Luttinger\cite{Luttinger64} (see also Refs.~\onlinecite{Shastry09}), the responses to $\delta T$ and $T\eta$ may be identified. Namely, when considering the response to the gravitational potential, which substitutes the temperature gradient, one should replace $\eta(\bfq,\omega)\rightarrow \delta T(\bfq,\omega)/T$.
For the purpose of finding the thermal conductivity from the dynamical heat density-heat density correlation function,
it is important that the limit $q\rightarrow 0$ should be taken before $\omega\rightarrow 0$. The heat current $\bfj_k(\bfq,\omega)$ may be found as a response to the gravitation potential $\eta(\bfq,\omega)$ and, furthermore, the static thermal conductivity $\kappa$ will be defined as the real part of the coefficient relating the heat current and $-\nabla T$ using the discussed relation between $\eta$ and $\delta T$. Eventually, the thermal conductivity $\kappa$ has to be extracted from the disorder-averaged heat density-heat density correlation function. As usual, translational invariance results from the averaging over disorder: $\langle\chi_{kk}(x_1,x_2)\rangle_{dis}=\chi_{kk}(x_{1}-x_2)$. Now, one may introduce the Fourier transform of the correlation function, and obtain $\kappa$ as follows \cite{Castellani87}
\be
\kappa =-\frac{1}{T}\lim_{\omega\rightarrow 0}\left(\lim_{q\rightarrow 0}\left[\frac{\omega}{\bfq^2} \mbox{Im}\chi_{kk}(\bfq,\omega)\right]\right)\label{eq:kappa}.
\ee
The calculation of the thermal conductivity in this paper will be based on this formula.

The correlation function $\chi_{kk}$ obeys the following two important relations
\be
&&\chi_{kk}(\bfq=0,\omega\rightarrow 0)=0,\label{eq:q=0}\label{eq:twolimit}\\
&&\chi_{kk}(\bfq\rightarrow 0,\omega=0)=-c_\mu T.\;\label{eq:twolimits}
\ee
Eq.~\eqref{eq:twolimit} reflects the conservation laws of energy and particle number, while Eq.~\eqref{eq:twolimits} relates the static part of the correlation function to the specific heat $c_\mu$ per unit volume at constant chemical potential.

\section{Heat density and Coulomb interaction in the extended NL$\sigma$M}
\label{sec:CoulombNLSM}

The definition of the heat density and the associated heat current for the electron gas has been at the center of a controversial discussion in recent works on thermal transport. Since this question is of fundamental importance for the calculation of the thermal conductivity, we will devote special attention to it. In view of the relation $k=h-\mu n$, and since the expressions for the particle density and the particle current are well-known, the mentioned discussion focuses around the definition of the energy (or hamiltonian) density and the energy current. At first sight, the answer seems straightforward, as one can construct the energy-momentum tensor (EMT) for the system of interacting electrons in a canonical way. Knowledge of the EMT allows to read off the continuity equation relating the energy density and energy current. Two problems arise in this context:

1)~The continuity equation obtained from the EMT relates a three-dimensional energy density to a three-dimensional energy current, while the problem of thermal transport for the two-dimensional electron gas requires knowledge of two-dimensional densities and currents.

2)~The canonical EMT is not gauge invariant.

The first point will be addressed in this section, where we suggest a simple procedure to project the three-dimensional quantities onto the plane. The second point, the problem of gauge-invariance, will be addressed in Appendix~\ref{sec:belinfante}, where we remind the reader of the field-theoretical construction of the so-called Belinfante EMT,\cite{Belinfante40,Greiner96} which results in a gauge-invariant expressions for the energy density and current. We would like to remark in this context that these expressions have already been obtained in Ref.~\onlinecite{Catelani05}, appendix B without making a connection with the Belinfante tensor.

\subsection{Projection of the electric field onto the charge carrying plane}
\label{sec:Efieldprojection}

Before discussing the derivation of the extended NL$\sigma$M with gravitational potentials, we would like to describe the main elements of the projection procedure separately. To this end, we will employ the following notation for spatial vectors: $\bfr=(x,y)^T$ is a $2d$ vector, $\underline{\bfr}=(x,y,z)^T$ is a $3d$ vector, and $\bfr_\circ=(x,y,0)^T$ denotes $\bfr$ embedded into the $3d$ space. We will assume that the $2d$ electron gas (2DEG) is located in the $xy$-plane, while the $z$-direction is perpendicular to this plane. We will also use the notation $x=(\bfr,t)$ for a combination of the $2d$ vector and time; for example, the two-dimensional number density is $n(x)={\psi}^\dagger_x\psi_x$.

The three dimensional hamiltonian density consists of a non-interacting and an interacting part $\underline{h}=\underline{h}_0+\underline{h}_{int}$. The transition to the two-dimensional density is straightforward for $\underline{h}_0$. We focus our attention on the interaction part. In the Coulomb gauge, it is given as (see formula~\eqref{eq:T00final} of Appendix~\ref{sec:belinfante})
\be
\underline{h}_{int}(\underline{\bfr},t)=\frac{1}{8\pi}\left[\underline{\bf E}^\parallel(\underline{\bfr},t)\right]^2,\label{eq:underlineh}
\ee
where $\underline{\bf E}^\parallel(\underline{\bfr},t)=-\underline{\nabla} \varphi(\underline{\bfr},t)$ and
\be
\varphi(\underline{\bfr},t)&=&\int d\bfr' \frac{en(\bfr',t)}{|\underline{\bf r}-\bfr'_{\circ}|}.
\ee
Here, $\underline{n}(\underline{\bfr},t)=n(\bfr,t)\delta(z)$, and $n$ denotes the $2d$ density of electrons confined to the $2d$ plane. Clearly, the field ${\bf E}^\parallel$ is non-zero outside of the 2DEG. In order to obtain a two-dimensional energy density, we integrate in the perpendicular coordinate $z$ as
\be
h_{int}(x)=\int dz \;\underline{h}_{int}(\underline{\bfr},t)\label{eq:projection}
\ee
It is instructive to transform the interaction term
\be
\left[\underline{\bf E}^\parallel(\underline{\bfr},t)\right]^2
=-\varphi(\underline{\bfr},t)\underline{\nabla}^2\varphi(\underline{\bfr},t)+\frac{1}{2}\underline{\nabla}^2\varphi^2(\underline{\bfr},t).
\ee
Using the Poisson equation $-\underline{\nabla}^2\varphi(\underline{\bf r},t)=4\pi en(\bfr,t)\delta(z)$, this decomposition allows us to write
\be
{h}_{int}(x)&=&\frac{1}{2}\int_{\bfr'} n(\bfr,t) V_0(\bfr_\circ-\bfr'_\circ) n(\bfr',t)\label{eq:hpro}\\
&+&\frac{1}{16\pi e^2}\nabla^2\int dz\left[\int_{\bfr'}V_0(\underline{\bfr}-\bfr_\circ')n(\bfr',t)\right]^2.\no
\ee
As a consequence of this integration in $z$, Eq.~\eqref{eq:projection}, $\nabla^2$ appears in the second term instead of the original $\underline{\nabla}^2$. It is clear now that interaction part of the Hamiltonian is recovered from $h_{int}$ by an integration over the 2d plane.
\be
H_{int}&=&\int d\bfr \;h_{int}(x)\no\\
&=&\frac{1}{2}\int_{\bfr,\bfr'}\; n(\bfr,t)\;V_{0}(\bfr_\circ-\bfr'_\circ)\;n(\bfr',t),
\ee
where $V_0(\underline{\bfr})=e^2/|\underline{\bfr}|$ is the familiar Coulomb interaction term.

Returning to Eq.~\eqref{eq:hpro}, we note that the first term can (loosely) be interpreted as a projection of the electric field onto the charge it originates from. The second term is a correction, for which the point of observation does not coincide with the position of the charge. Later, it will be shown that the second term on the right hand side of Eq.~\eqref{eq:hpro} does not contribute to the correlation function in the long wavelength limit due to the presence of $\nabla^2$. The crucial point here is that the interaction potential $V_0$ becomes screened due to the conducting plane. For the screened potential, unlike for the bare $V_0$, one can neglect the second term in Eq.~\eqref{eq:hpro} in the limit of small gradients.

\subsection{Fermionic action with gravitational potentials}

In this section, we prepare the derivation of the NL$\sigma$M by introducing the gravitational potential into the action and further by decoupling the interaction term. Let us recall that according to the discussion in the previous section the full expression for the two-dimensional hamiltonian density is $h=h_0+h_{int}$, where $h_{int}$ is given in Eqs.~\eqref{eq:underlineh} and \eqref{eq:projection} and $h_0$ describes propagation of particles in the presence of disorder
\be
h_{0}(x)&=&\frac{1}{2m}\nabla{\psi}^\dagger_x\nabla{\psi}_x+u_{dis}(\bfr)n(x)\label{eq:h0}.
\ee

In order to write the action in the presence of the gravitational potentials in a compact form, it is convenient to define a matrix $\hat{\eta}'$ acting in the space of fields $\vec{\psi}=(\psi_+,\psi_-)^T$ as
\be
\hat{\eta}'=\left(\ba{cc}\eta_1+\eta_2&0\\0&\eta_1-\eta_2\ea\right).
\ee

Then, the action $S$ defined on the Keldysh contour can be written as
\be
S[\vec{\psi}^\dagger,\vec{\psi},\hat{\eta}']
&=&\int_x\; \vec{\psi}^\dagger\left(i\partial_t-[u_{dis}-\mu](1+\hat{\eta}')\right)\hat{\sigma}_3\vec{\psi}\no\\
&&-\int_x\frac{1}{2m}\nabla\vec{\psi}^\dagger(1+\hat{\eta}')\hat{\sigma}_3\nabla \vec{\psi}\no\\
&&-\frac{1}{8\pi}\int_t \int d\underline{\bfr} \;\vec{{\bf E'}}^T(1+\hat{\eta}')\hat{\sigma}_3\vec{\bf E'}
\ee
Here, and in the following, we write $\int_t=\int_{-\infty}^\infty dt$ and $\int_{x}=\int_{\bfr,t}$. Summation over the spin degrees of freedom is implicit. The third Pauli matrix $\hat{\sigma}_3$ acts in the space of forward and backward fields. From now on, all matrices acting in the Keldysh space will indicated by a hat.
Besides, we wrote $\vec{\bf E'}=({\bf E'}^\parallel_+,{\bf E'}^\parallel_-)^T$, where
\be
e{\bf E'}^\parallel_\pm(\underline{\bf r},t)=-\underline{\nabla}\int d\bfr' V_0(\underline{\bf r}-\bfr'_\circ)n_{\pm}(\bfr',t).
\ee

Since our strategy is to project the entire problem onto the conducting plane, it will be assumed that $\eta'=\eta'(x)$ does not depend on $z$. Note that as a result, $\eta_2(x)$ couples to the two-dimensional heat density. The hamiltonian part of this $2d$ heat density corresponds to the one introduced in Eqs.~\eqref{eq:projection} and \eqref{eq:h0}.

Next, the Keldysh rotation can be performed.\cite{Larkin75,Kamenev11} To this end, we introduce new fermionic fields
\be
\vec{\Psi}^\dagger=\vec{\psi}^\dagger \hat{L}^{-1},\quad\vec{\Psi}=\hat{L}\hat{\sigma}_3\vec{\psi},\quad \hat{L}=\frac{1}{\sqrt{2}}\left(\ba{cc}1&-1\\1&1\ea\right).\label{eq:trafo}
\ee
With the help of the two matrices $\hat{\gamma}_1=\hat{\sigma}_0$, $\hat{\gamma}_2=\hat{\sigma}_1$ in Keldysh space, one may form the matrix of gravitational potentials $\hat{\eta}=\sum_{k=1,2}\eta_k\hat{\gamma}_k$. The action after this rotation reads
\be
S[\vec{\Psi}^\dagger,\vec{\Psi},\hat{\eta}]
&=&\int_x \;\vec{\Psi}^\dagger \left(i\partial_t-[u_{dis}-\mu](1+\hat{\eta})\right)\vec{\Psi}\no\\
&&-\int_x\frac{1}{2m^*}\nabla \vec{\Psi}^\dagger(1+\hat{\eta})\nabla\vec{\Psi}\no\\
&&-\frac{1}{16\pi}\int_t \int d\underline{\bfr} \;\vec{\bf E}^T(1+\hat{\eta})\hat{\gamma}_2\vec{\bf E},\label{eq:SKeld}
\ee
where
\be
e\vec{\bf E}_k(\underline{\bfr},t)=-\nabla\int d\bfr'\; V_0(\underline{\bfr}-\bfr'_\circ)\Psi(\bfr',t)\hat{\gamma}_k\Psi(\bfr',t).
\ee

The last term in Eq.~\eqref{eq:SKeld} contains four fermionic fields. We introduce two real Hubbard-Stratonovich fields $\theta_{1,2}$, forming the matrix $\hat{\theta}=\sum_{k=1,2}\theta_k\hat{\gamma}_k$ to decouple this term. Note that in the case of the Fermi liquid, in order to decouple all interaction terms, four Hubbard-Stratonovich matrix fields $\hat{\theta}^l$ have to be introduced, where the index $l=0-3$ denotes the density and spin density interaction channels. For the Coulomb problem, without account of Fermi liquid-type interactions, only the singlet channel, $l=0$, is involved. For this reason, no index $l$ will be used here. After these transformations, the partition function $\mathcal{Z}=\int D[\hat{\theta}]D[\vec{\Psi}^\dagger,\vec{\Psi}]\exp(iS[\vec{\Psi}^\dagger,\vec{\Psi},\hat{\theta}])$ can be written with the use of action
\begin{align}
&S[\vec{\Psi}^\dagger,\vec{\Psi},\hat{\theta}]=\int_x \;\vec{\Psi}^\dagger \left(i\partial_t-[u_{dis}-\mu](1+\hat{\eta})+\hat{\theta}\right)\vec{\Psi}\no\\
&-\int_x\frac{1}{2m^*}\nabla \vec{\Psi}^\dagger(1+\hat{\eta})\nabla\vec{\Psi}+\int_{x,x'} \vec{\theta}^T \hat{\mathcal{V}}^{-1}_\eta \hat{\gamma}_2\vec{\theta},\label{eq:SV}
\end{align}
where
\be
\hat{\mathcal{V}}^{-1}_\eta(x,x')&=&\int d\underline{\bfr}''d\underline{\bfr}'''V_0^{-1}(\bfr_\circ-\underline{\bfr}'')\no\\
&&\times \hat{V}_\eta(\underline{\bfr}'',\underline{\bfr}''',t,t')V_0^{-1}(\underline{\bfr}'''-\bfr'_\circ).
\ee
Here, $\hat{V}_\eta$ fulfils the generalized Poisson equation
\begin{align}
-\underline{\nabla}((1+\hat{\eta}(x))\underline{\nabla})\hat{V}_\eta(\underline{\bfr},\underline{\bfr}',t,t')=4\pi e^2\delta(\underline{\bfr}-\underline{\bfr}')\delta(t-t').
\end{align}

A useful relation can be obtained for the electron interaction $\hat{\mathcal{V_{\eta}}}$ in the action $S$ in Eq.~\eqref{eq:SV}
\begin{align}
\hat{\mathcal{V}}_\eta=V_0\hat{V}^{-1}_{\eta}V_0
=\frac{1}{2}\{1+\hat{\eta},V_0\}+\frac{1}{8\pi e^2}V_0(\nabla^2\hat{\eta}) V_0.\label{eq:useful}
\end{align}
For the sake of simplicity, we used a matrix notation for the spatial coordinates here. As one can see, the above expression reproduces the interaction term given in \eqref{eq:hpro}. Note that the expression above is not an approximation; there are no higher order terms in $\eta$. Naturally, $\hat{\mathcal{V}}_{\eta=0}(x,x')=V_0(\bfr_\circ-\bfr'_\circ)\delta(t-t')$, making contact with the theory of the two-dimensional electron liquid in the absence of the gravitational potential.
Since $\eta_2(x)$ couples to the heat density, the action given in \eqref{eq:SV} taken together with the relation \eqref{eq:useful} reflects the form of the hamiltonian density stated by Eqs.~\eqref{eq:h0} and \eqref{eq:hpro}.

\subsection{The extended non-linear sigma model}
\label{sec:extendedNLSM}

In this paper, we concentrate on peculiarities of thermal transport related to the Coulomb interaction. A compact description of our approach to the analysis of heat transport in a disordered Fermi liquid with short-range interactions can be found in Ref.~\onlinecite{Schwiete14a}, while a detailed discussion of the NL$\sigma$M extended by the gravitational potentials was presented in Ref.~\onlinecite{Schwiete14b}.

As it has been explained in the Introduction, we are interested in small energies and long distances. For distances exceeding the mean free path, the physics is described by slow diffusion modes (i.e., modes describing density relaxation in the presence of disorder) rather than single-particle excitations. Therefore, the fermionic fields $\psi$ and $\psi^\dagger$ have to be integrated out. Furthermore, averaging over disorder realizations can be performed assuming that disorder is weak in the sense that $\eps_F\tau\gg 1$, where $\eps_F$ is the Fermi energy and $\tau$ the transport scattering time. Then, the entire physics of the diffusion modes (the so-called diffusons) can be encoded in the fluctuations of a matrix $\underline{\hat{Q}}_{t,t'}(\bfr)$ with respect to its saddle point position $\underline{\hat{Q}_0}(\bfr,t,t')=\hat{\Lambda}_{t-t'}$, where
\be
\hat{\Lambda}_{\eps}=\left(\ba{cc} 1&2\mathcal{F}_\eps\\0&-1\ea\right)=\hat{u}_\eps\hat{\sigma}_3\hat{u}_\eps,\quad \hat{u}_\eps=\left(\ba{cc}1&\mathcal{F}_\eps\\0&-1\ea\right)
\ee
and $\mathcal{F}_\eps=\tanh\left({\eps}/{2T}\right)$ is the fermionic equilibrium distribution function.
Here and elsewhere below, $2\times 2$ matrices denoted by the hat symbol act in Keldysh space, with the rotation $\hat{L}$ being already performed. The manifold of low-lying gapless excitations is described by rotations
\be
\underline{\hat{Q}}=\hat{u}\circ \hat{Q}\circ \hat{u},\quad  \hat{Q}=\hat{U}\circ\hat{\sigma}_3\circ \hat{\overline{U}},
\ee
where $\hat{U}=\hat{U}_{t,t'}(\bfr)$, and $(\hat{U}\circ \hat{\overline{U}})_{t,t'}=\delta(t-t')$. The $\circ$-symbol denotes a convolution in time.

It will be convenient to release the disorder term $u_{dis}$ in the action $S$ from the explicit dependence on the gravitational potentials. To this purpose, the transformation ("$\lambda$-transformation")
\be
\vec{\psi}\rightarrow \hat{\lambda}^\frac{1}{2}\vec{\psi}, \qquad \vec{\psi}^\dagger\rightarrow \vec{\psi}^\dagger \hat{\lambda}^\frac{1}{2},\qquad \hat{\lambda}=(1+\hat{\eta})^{-1}
\ee
of the fermionic fields was implemented in Refs.~\onlinecite{Schwiete14a,Schwiete14b}. For details of the $\lambda$-transformation, we refer to these papers.
As a result of the $\lambda$-transformation, $\hat{\eta}$ appears through the matrix $\hat{\lambda}=1-\hat{\gamma}_1\eta_1-\hat{\gamma}_2\eta_2+2
\hat{\gamma}_2\eta_1\eta_2+\dots$. For the calculation of the correlation function according to Eq.~\eqref{eq:chisources}, one needs to consider the expansion of $\hat{\lambda}$ up to second order in $\hat{\eta}$. For the dynamical part of the correlation function, however, only the terms linear in $\hat{\eta}$ are required.

Starting from the fermionic action displayed in Eq.~\eqref{eq:SV}, one may apply the $\lambda$-transformation and subsequently follow the traditional route to derive the NL$\sigma$M suitable for description of disordered electrons interacting via the Coulomb interaction. When written in terms of deviations of the matrix field $\underline{Q}$ from its saddle point, $\underline{\delta Q}=\underline{\hat{Q}}-\hat{\Lambda}$, the model looks as follows
\begin{align}
S&=\frac{\pi\nu_0 i}{4}\Tr\left[ D(\nabla \hat{Q})^2+2i \{\hat{\eps},\hat{\lambda}\} \underline{\delta \hat{Q}} \right]\no\\
&-\frac{\pi^2\nu^2}{4}\int_{xx'} \tr[\hat{\lambda}\hat{\gamma}_i\underline{\delta \hat{Q}_{tt}}(\bfr)](\hat{\gamma}_2\hat{\mathcal{V}}^{s}_{\eta}(x,x'))_{ij} \no\\
&\times\tr[\hat{\lambda}\hat{\gamma}_j \underline{\delta \hat{Q}_{t't'}}(\bfr')]\no\\
&+Tc_0 \int_x\vec{\eta}^T(x)\hat{\gamma}_2\vec{\eta}(x).
\label{eq:actioneta}
\end{align}
Here, the $\tr$-symbol includes a trace in Keldysh space, an integration over
frequencies (when the matrix $Q$ is written in frequency space), and a summation over spin degrees of freedom; the symbol $\Tr$ includes, in addition, an integration over coordinates. The first two terms in Eq.~\eqref{eq:actioneta} describe diffusion in the absence of the electron-electron interaction; $D$ is the diffusion coefficient; $\nu_0$ is the single particle density of states per spin direction. The electron-electron interaction acts only in the singlet channel (no Pauli matrices acting in the spin space are present) as it should be for the Coulomb interaction. The term in the last line describes the contributions to the static part of the heat-density heat-density correlation function originating from fermionic degrees of freedom (i.e., without participation of the diffusion modes); $c_{0}=2\pi^2\nu_0 T/3$ is the specific heat of electrons. We suppressed an additional term that is linear in $\eta_2$ and required only for the calculation of the heat density itself.

The Coulomb interaction entering the action $S$ is statically screened, $\hat{\mathcal{V}}_\eta^{s}=(\hat{\mathcal{V}}^{-1}_\eta+2\nu_0\hat{\lambda})^{-1}$. This formula is symbolical: Both $\hat{\mathcal{V}}_\eta^{s}$ and $\hat{\mathcal{V}}_\eta$ depend on three-dimensional spatial coordinates, but screening takes place in the two-dimensional plane. Importantly, $\hat{\lambda}$ appears in the term responsible for screening. At zeroth order in $\eta$, the interaction $\mathcal{V}_\eta^{s}$ coincides with the statically screened Coulomb interaction $\mathcal{V}_{\eta=0}^{s}\equiv V_0^{s}=(V_0^{-1}+2\nu_0)^{-1}$, where again screening occurs in the plane only. The relation \eqref{eq:useful} allows us to obtain a regular expansion for $\hat{\mathcal{V}}_\eta^s$ in powers of $\eta$
\be
\hat{\mathcal{V}}_\eta^s=\frac{1}{2}\{1+\hat{\eta},V_0^s\}+\frac{1}{8\pi e^2}V_0^s(\nabla^2\hat{\eta})V_0^s+\mathcal{O}(\eta^2).\label{eq:Vetas}
\ee
Since $V_0^s$ is not singular anymore, one can neglect in $\hat{\mathcal{V}}_\eta^s$ the second term on the right hand side in the limit of small gradients. Thus, owing to screening, the point of observation coincides with the position of the charges when finding the heat density correlation function of a system of conducting electrons confined within a $2d$ plane. (In the case of the bare, i.e., unscreened, Coulomb interaction one cannot neglect the second term in $\hat{\mathcal{V}}_{\eta}$ even in the limit of small gradients.) As a consequence, all subsequent considerations involve the effective two-dimensional Coulomb interaction with
$V_0^s={2\pi e^2}/(\bf |q|+\kappa_s)$, where $\kappa_s=4\pi e^2 \nu_0$ is the inverse screening radius.

\section{Dynamical parts of the correlation functions - General formulas}
\label{sec:Dynamic}

\subsection{Dynamical part of the heat density correlation function}

Here, we focus on the dynamical part of the correlation function $\chi_{kk}^{dyn}$, for which the corresponding diagrams are reducible with respect to cutting a single diffuson. The starting point for all subsequent calculations will be the Keldysh NL$\sigma$M action in the presence of the gravitational potentials, Eq.~\eqref{eq:actioneta}. Only the $Q$-dependent part of the action \eqref{eq:actioneta} is relevant for the calculation (the last term in $S$ can be abandoned). In addition, we may restrict ourselves to terms of linear order in $\hat{\eta}$ in the action. This allows us, in particular, to use the linear approximation for the interaction $\hat{\mathcal{V}}_\eta^s$ displayed in Eq.~\eqref{eq:Vetas}. To linear order in $\hat{\eta}$, the $Q$-dependent part of the action reads
\begin{align}
S_{lin}&=\frac{\pi\nu_0 i}{4}\Tr\left[ D(\nabla \hat{Q})^2+2i \{\hat{\eps},1-\hat{\eta}\} \underline{\delta \hat{Q}} \right]\\
&-\frac{\pi^2\nu_0^2}{4}\int_{\bfr\bfr',t} \tr[(1-\hat{\eta})\hat{\gamma}_i\underline{\delta \hat{Q}_{tt}}(\bfr)]\hat{\gamma}_2^{ij}V^{s}_{0}(\bfr-\bfr') \no\\
&\times\tr[\hat{\gamma}_j \underline{\delta \hat{Q}_{tt}}(\bfr')].\no
\end{align}
We decompose
\be
S_{lin}=S_{\eta=0}+S_{\eps\eta}+S_{\eta V}.
\ee
with two types of source terms in the action. The first one is already present in the noninteracting theory
\begin{align}
S_{\eta\eps}=&\frac{\pi\nu_0}{2}\Tr[\{\hat{\eps},\hat{\eta}\}\underline{\delta \hat{Q}}].\label{eq:Setaeps}
\end{align}
The other source term is specific for the interacting problem
\begin{align}
S_{\eta V}
=&\frac{(\pi\nu_0)^2}{4}\int_{\bfr\bfr',t}\tr[\hat{\eta}\hat{\gamma}_i\underline{\delta \hat{Q}_{tt}}(\bfr)]\label{eq:SetaV}\\
&\times \hat{\gamma}_2^{ij}V^{s}_{0}(\bfr-\bfr') \tr[\hat{\gamma}_j \underline{\delta \hat{Q}_{tt}}(\bfr')].\no
\end{align}
As will become clear below, the existence of this vertex is of crucial importance for the internal consistency of the theory, in particular with respect to the conservation of energy.

\begin{figure}
\includegraphics[width=5cm]{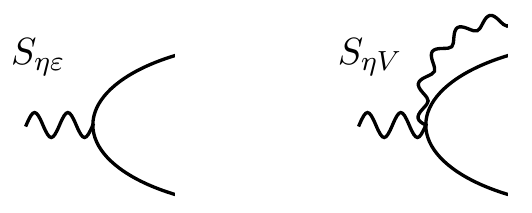}
\caption{The frequency and interaction vertices originating from the source terms $S_{\eta\eps}$ and $S_{\eta V}$ as introduced in Eqs.~\eqref{eq:Setaeps} and \eqref{eq:SetaV}, respectively. The density vertex arising in connection with the calculation of the density-density correlation function will be drawn in the same way as the frequency vertex.}
\label{fig:TwoVertices}
\end{figure}

The two source terms of Eq.~\eqref{eq:Setaeps} and Eq.~\eqref{eq:SetaV} give rise to two vertices in the diagrammatic representation, which we will refer to as the frequency vertex and the interaction vertex, respectively. They are displayed in Fig.~\ref{fig:TwoVertices}. One can further distinguish between vertices originating from a differentiation with respect to $\eta_2$ and $\eta_1$. For the sake of definiteness, we will draw the vertices related to $\eta_2$ on the left hand side and those related to $\eta_1$ on the right hand side of a diagram.

Correspondingly, for finding the dynamical part, we need to calculate
\begin{align}
\chi^{dyn}_{\eps \eps}(x_1,x_2)&=-\frac{i}{2}(\pi\nu_0)^2\int_{\eps_i}\mbox{e}^{-it_1(\eps_1-\eps_2)+it_2(\eps_3-\eps_4)}\label{eq:chiepseps}
\\&\times \langle\overline{\eps}_{12}\tr[\hat{\gamma}_2\underline{\delta \hat{Q}_{\eps_1\eps_2}}(\bfr_1)]\overline{\eps}_{34}\tr[\hat{\gamma}_1\underline{\delta \hat{Q}_{\eps_4\eps_3}}(\bfr_2)]\rangle_r,\no
\end{align}
where $\eps_{ij}=(\eps_i+\eps_j)/2$, together with the term
\begin{align}
&\chi^{dyn}_{\eps V}(x_1,x_2)\\
=&-\frac{i}{8}(\pi\nu_0)^3\int_{\bfr_3,\eps_i}\mbox{e}^{-it_1(\eps_1-\eps_2)}\langle\overline{\eps}_{12}\tr[\hat{\gamma}_2\underline{\delta \hat{Q}_{\eps_1\eps_2}}(\bfr_1)]\no\\
&\times\tr[\hat{\gamma}_1\hat{\gamma}_{i}\underline{\delta \hat{Q}_{t_2t_2}}(\bfr_2)]\hat{\gamma}_2^{ij}V_0^s(\bfr_2-\bfr_3)\tr[\hat{\gamma}_j\underline{\delta \hat{Q}_{t_2t_2}}(\bfr_3)]\rangle_r\no
\end{align}
and the analogous term $\chi^{dyn}_{V\eps}(x_1,x_2)$. We introduced the notation $\int_\eps=\int \frac{d\eps}{2\pi}$. The index $r$ in these formulas indicates that only those contributions should be selected that are reducible with respect to a single diffuson. Averaging $\left\langle\dots\right\rangle$ is with respect to the action $S_{\eta=0}$.  A term with two interaction vertices exists, $\chi_{VV}$, but is not written because it does not contribute to the dynamical part of the correlation function in the one-loop approximation, but only to the static part.

\subsection{The dynamical part of the density-density correlation function}
It is instructive to compare the calculation of the heat density-heat density correlation function to that of the density-density correlation function in the same formalism. This correlation function can generated from the source term $S_\varphi=\pi\nu_0 \Tr[\hat{\varphi} \underline{\hat{Q}}]$ by differentiation
\be
\chi_{nn}(x_1,x_2)=\left.\frac{i}{2}\frac{\delta^2 \mathcal{Z}}{\delta\varphi_2(x_1)\delta\varphi_1(x_2)}\right|_{\eta_2=\eta_1=0},\label{eq:chinn}
\ee
in analogy to Eq.~\eqref{eq:chisources}. From this formula, one obtains the expression
\be
&\chi^{dyn}_{nn}(x_1,x_2)=-\frac{i}{2}(\pi\nu_0)^2\int_{\eps_i}\mbox{e}^{-it_1(\eps_1-\eps_2)+it_2(\eps_3-\eps_4)}\no
\\&\times \langle\tr[\hat{\gamma}_2\underline{\delta \hat{Q}_{\eps_1\eps_2}}(\bfr_1)]\tr[\hat{\gamma}_1\underline{\delta \hat{Q}_{\eps_4\eps_3}}(\bfr_2)]\rangle_r.
\ee
In contrast to the heat transport, only a single vertex exists, the density vertex. For this vertex, we will use the same graphical representation as for the frequency vertex, i.e., the one displayed in the left part of Fig.~\ref{fig:TwoVertices}.

For the dynamical part of the density-density correlation function rescattering on the short-range part of the electron-electron interaction is allowed, while for the heat density correlation function this is impossible. In spite of these differences, within the RG-interval of energies the WFL holds; see Refs.~\onlinecite{Schwiete14a,Schwiete14b,Castellani87}. This indicates that there are non-trivial but robust connections between the interaction vertices and the interaction amplitudes that are fulfilled for the Fermi liquid and remain valid even during the course of the RG transformations.

\subsection{Perturbation theory and dynamical screening}

\begin{figure}
\includegraphics[width=7cm]{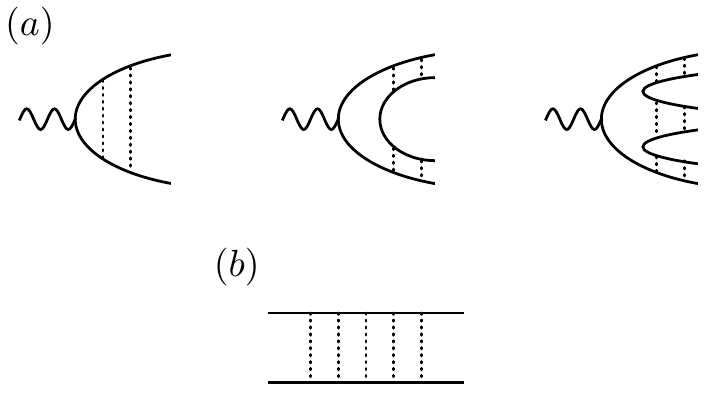}
\caption{(a) Diagrammatic representation of the frequency vertex in the sigma model. The first three terms of the expansion in $P$-modes are displayed. (b) Diffuson.}
\label{fig:VertexExpansion}
\end{figure}

In the calculation of the correlation functions, an expansion of $\hat{Q}$ in deviations from the metallic saddle point $\hat{\sigma}_3$ is needed. For the sake of definiteness, we choose the exponential parametrization $\hat{U}=\exp(-\hat{P}/2)$ with $\{\hat{\sigma}_3,\hat{P}\}=0$, so that $\hat{Q}=\hat{\sigma}_3\exp(\hat{P})$. Fortunately, an expansion to low orders in the generator $\hat{P}$ is sufficient for our calculation. As an example, the diagrammatic representation of the frequency vertex in the sigma model is illustrated in Fig.~\ref{fig:VertexExpansion} (a); the interaction vertex is represented analogously.
When the expansion of $\hat{Q}$ in $\hat{P}$ is implemented in $S_{\eta=0}$, this gives among other terms rise to the quadratic action $S_0$ of the noninteracting theory
\begin{align}
S_0=-\frac{i\pi\nu_0}{4}\Tr\left[D(\nabla \hat{P})^2-2i\hat{\eps}\hat{\sigma_3}\hat{P}^2\right].\label{eq:S0}
\end{align}
The action $S_0$ describes the propagation of diffusion modes, so-called "diffusons" [see Fig.~\ref{fig:VertexExpansion} (b)]
\be
\mathcal{D}_{\bfq,\omega}=\frac{1}{D\bfq^2-i\omega}.\label{eq:diffuson}
\ee
Note that $\mathcal{D}$ is the retarded diffuson, the advanced diffuson will be denoted as $\overline{\mathcal{D}}$ and is related to the retarded diffusion as $\overline{\mathcal{D}}_{\bfq,\omega}=\mathcal{D}_{\bfq,-\omega}$. In the perturbative calculations involving diffusion modes, $S_0$ given by Eq.~\eqref{eq:S0} serves as a starting action. The necessary contractions for the Gaussian averaging can be performed with the help of the contraction rules that we formulate in Appendix~\ref{app:Contractions}.

A distinctive feature of the sigma model for interacting systems is that upon expansion of $S_{\eta=0}$ in deviations from the saddle point the interaction potential contributes to the quadratic form in the $P$-modes. This allows one to incorporate Fermi liquid effects into the propagation of diffusion modes in an automatic way. These effects can be interpreted as rescattering of diffusons by the electron-electron interaction, or alternatively as a modification of the interaction amplitudes by diffusons. An example of such a processes is presented in Fig.~\ref{fig:nncorrelation}, which represents the dynamic part of the polarization operator.
[A discussion of the diffuson propagator modified by the electron interactions in the Keldysh formalism can be found in Sec.~III B of Ref.~\onlinecite{Schwiete14}.] In the present context it is more convenient to delegate the output of %non-perturbative
the resummation to the interaction itself. As a result, the dynamically screened Coulomb interaction should be used instead of the statically screened $V_0^s$.

In the Keldysh formalism, the dynamically screened interaction acquires a non-trivial matrix structure in Keldysh space:
\begin{align}
\hat{V}^{ij}_{\bfk,\nu}=\left(\ba{cc} V^K_{\bfk,\nu}&V^R_{\bfk,\nu}\\V^A_{\bfk,\nu}&0\ea\right),\quad V^R_{\bfk,\nu}=\frac{1}{V_0^{-1}(\bfk)+\mathcal{P}^R_{\bfk,\nu}}.
\end{align}
In this formula, $V_0(\bfk)=2\pi e^2/|\bfk|$ is the effective two-dimensional Coulomb interaction and $\mathcal{P}^R(\bfk,\nu)=2\nu_0 {D\bfk^2}/(D\bfk^2-i\nu)$ is the retarded polarization operator. The advanced and Keldysh components of $\hat{V}$ are defined as $V^A_{\bfk,\nu}=V^R_{\bfk,-\nu}$ and $V^K_{\bfk,\nu}=\mathcal{B}_{\nu}(V^R_{\bfk,\nu}-V^A_{\bfk,\nu})$, where $\mathcal{B}_\nu=\coth(\nu/2T)$ is the bosonic distribution function.

\section{Structure of the correlation functions}
\label{sec:gstructure}
Before turning to the calculation based on the specific formalism used in this paper, it is instructive to discuss the general structure of the heat density-heat density correlation function $\chi_{kk}(\bfq,\omega)$. In particular, we are interested in the constraint given by Eq.~\eqref{eq:twolimit}, which is a consequence of the fact that $\chi_{kk}(\bfq,\omega)$ describes the propagation of the heat density under the condition when the entropy is a conserved quantity.

We are interested in the singular behavior of $\chi_{kk}(\bfq,\omega)$ which depends on the order of taking the limits $q\rightarrow 0$ and $\omega\rightarrow 0$. We will assume in this section that all intermediate integrations have already been performed, and, correspondingly, all corrections arising from the RG-interval \emph{and} sub-temperature energy range have been absorbed into the constants which determine the correlation function. In other words, we will discuss the "ultimate" stage when everything that does not depend singularly on $\bfq$ and $\omega$ can be substituted by a constant. The remaining singular behavior originates from the diffusion propagation of electron-hole pairs, which for free electrons is described by the propagator
\be
\mathcal{D}(\bfq,\omega)=\frac{1}{D\bfq^2-i\omega}.\label{eq:Dfree}
\ee
In the presence of the electron interaction, this propagator has to be modified as will be described below.

In order to allow for a direct comparison with the density-density correlation function, $\chi_{kk}$ will be structured in the same way as $\chi_{nn}$. [A discussion of the density-density correlation function can be found in Refs.~\onlinecite{Finkelstein83,Castellani84,DiCastro04,Finkelstein10, Schwiete14}. The heat density-heat density correlation function has been analyzed in Ref.~\onlinecite{Castellani87}. However, the scattering processes, which are the center of our interest here, have not been considered so far for $\chi_{kk}$.] In both cases, the correlation function can be split into static and dynamical parts. As we have already mentioned in Sec.~\ref{sec:model}, see Eqs.~\eqref{eq:twolimit} and \eqref{eq:twolimits}, the static parts are related to the corresponding thermodynamic quantities: the compressibility in the case of $\chi_{nn}$, and specific heat in the case of $\chi_{kk}$. The dynamical parts should cancel the static ones in the limit $\bfq=0,\omega\rightarrow 0$, which is the way the conservation laws for particle number and energy manifest themselves. Our goal is to demonstrate how this works for $\chi_{kk}$. We start, however, with $\chi_{nn}$ for which this procedure is well established.

\subsubsection{The density-density correlation function}
\label{sec:nn}

\begin{figure}
\includegraphics[width=8cm]{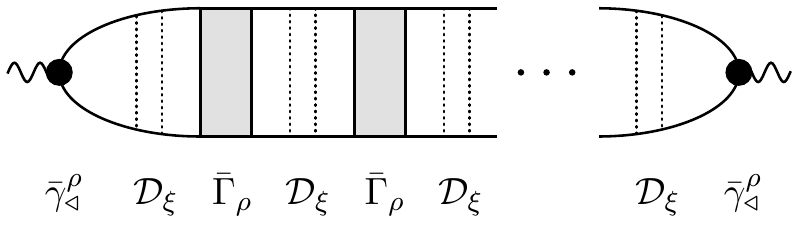}
\caption{The structure of the dynamical part of the density-density correlation function in accordance with Eq.~\eqref{eq:chinndyn}.}
\label{fig:nncorrelation}
\end{figure}

The density-density correlation function can be split into a static and a dynamical part
\be
\chi_{nn}(\bfq,\omega)=\chi_{nn}^{st}+\chi_{nn}^{dyn}(\bfq,\omega),\label{eq:chinndecomp}
\ee
where the static part is defined as $\chi_{nn}^{st}=\chi_{nn}(\bfq\rightarrow 0,\omega=0)$. Quite generally, the static and dynamical part can be further decomposed as follows
\be
\chi_{nn}^{st}&=&-2\nu_0\gamma_\bullet^\rho\\
\chi_{nn}^{dyn}(\bfq,\omega)&=&-2\nu_0 (\bar{\gamma}^\rho_\triangleleft)^2\frac{i\omega}{\mathcal{D}_\xi^{-1}(\bfq,\omega)+i\bar{\Gamma}_\rho\omega}.\label{eq:chinndyn}
\ee
Let us discuss the parameters appearing in the above expressions. %definition of $\chi^{dyn}_{nn}$.
As is well known, the static part of the correlation function is related to the compressibility $\chi_{nn}^{st}=-\partial n/\partial \mu$.
% and does not acquire quantum corrections.
Therefore,
\be
\gamma^\rho_\bullet=\frac{1}{2\nu_0}\frac{\partial n}{\partial\mu}.\label{eq:gammarhobullet}
\ee
The structure of the dynamical part of the correlation function is displayed in Fig.~\ref{fig:nncorrelation}.
The vertex corrections for the two scalar vertices are denoted by $\bar{\gamma}^\rho_\triangleleft$; $\bar{\Gamma}_\rho$ is the short range part of the singlet interaction amplitude. This means, in particular, that the long-range part of the Coulomb interaction is \emph{not} included in $\chi_{nn}$. The diffusion propagator modified by the electron interaction, the diffuson $\mathcal{D}_\xi$, is defined as
\be
\mathcal{D}_\xi(\bfq,\omega)=\frac{\xi^2}{D\bfq^2-iz\omega}.\label{eq:Dxi}
\ee
It incorporates the frequency renormalization $z$, introduced in Ref.~\onlinecite{Finkelstein83}, and the wave-function renormalization $\xi^2$. Using the relation $\chi_{nn}(\bfq=0,\omega\rightarrow 0)=0$, a direct consequence of particle number conservation, one deduces the following condition
\be
z_1=\frac{\xi^2(\bar{\gamma}_\triangleleft^\rho)^2}{\gamma_\bullet^\rho},\label{eq:z1}
\ee
where we use the notation $z_1$ for the combination $z_1=z-\xi^2\bar{\Gamma}_\rho$. Adding the static and the dynamical part, one finds in view of the Eqs.~\eqref{eq:z1} and Eq.~\eqref{eq:gammarhobullet} that
\be
\chi_{nn}(\bfq,\omega)=-\frac{\partial n}{\partial \mu}\frac{D\bfq^2}{D\bfq^2-iz_1\omega}.
\ee

The wave-function renormalization appears explicitly only in the diagrammatic approach. In the $\sigma$-model approach to the problem,~\cite{Finkelstein83, Schwiete14} one deals directly with the effective amplitude $\Gamma_\rho=\xi^2\bar{\Gamma}_\rho$ and the effective vertex correction $\gamma_\triangleleft^\rho=\xi \bar{\gamma}_\triangleleft^\rho$. 
Electric conductivity can be found from the relation
\be
\sigma=-e^2\lim_{\omega\rightarrow 0}\lim_{\bfq\rightarrow 0}\left[\frac{\omega}{\bfq^2}\mbox{Im}\chi_{nn}^R(\bfq,\omega)\right]
\ee
from which one deduces $\sigma=2\nu_0 e^2 D$.

Let us now turn to a specific model in which only the long-range part of the Coulomb interaction is accounted for. [Screening will be included, while the short-range Fermi liquid amplitudes both in the singlet and in the triplet channel are ignored. By contrast, the short-range amplitudes generated by the interplay of the Coulomb interaction and disorder have to be included.] It means that in the absence of the Fermi-liquid corrections, at the initial scale of the RG-integration, i.e., at $1/
\tau$, one has $\gamma_\bullet^\rho=\gamma_\triangleleft^\rho=\bar{\gamma}_\triangleleft^\rho=z_1=\xi^2=z=1$, as well as $\bar{\Gamma}_\rho=\Gamma_\rho=0$. From the previous analysis,\cite{Finkelstein83,Castellani84,DiCastro04,Finkelstein10} it is well understood that the compressibility $\chi_{nn}^{st}=-\partial n/\partial \mu$
does not acquire quantum corrections. In the explicit calculations presented below, we will take the relation $\gamma_\bullet^\rho=1$ for the static part for granted, and only analyze the dynamical part of the correlation function. The goal will be to identify the corrections $\delta \xi^2$, $\delta D$, $\delta z$, $\delta \bar{\Gamma}_\rho$, $\delta \bar{\gamma}_\triangleleft^\rho$ in the expression
\begin{align}
&\chi^{dyn}_{nn}(\bfq,\omega)\approx -2\nu_0\frac{i\omega (1+\delta \xi^2+2\delta\bar{\gamma}^\rho_{\triangleleft})}{(D+\delta D)\bfq^2-i\omega(1+\delta z-\delta\bar{\Gamma}_\rho)}\no\\
\approx& -2\nu_0 i\omega\mathcal{D}\no\\
&-2\nu_0 i\omega\left[\delta \xi^2(D\bfq^2-i\omega)-\delta D\bfq^2+i\omega(\delta z-\delta \bar{\Gamma}_\rho)\right]\mathcal{D}^2\no \\
&-2\nu_0 i\omega  (2\delta\bar{\gamma}^\rho_{\triangleleft})\mathcal{D}.\label{eq:nnmodel}
\end{align}

In particular, we need to check the relation $\delta z_1=0$, which implies
\be
\delta z=\delta (\xi^2\bar{\Gamma}_\rho)=\delta\bar{\Gamma}_\rho.\label{eq:check1}
\ee
The first equality follows directly from the definition of $z_1$, while the second equality is a consequence of the fact that initially $\bar{\Gamma}_\rho=0$ and $\xi^2=1$. A second constraint reads
\be
2\delta \gamma_\triangleleft^\rho=\delta \xi^2+2\delta \bar{\gamma}_\triangleleft^\rho=0.\label{eq:check2}
\ee
This constraint follows from Eq.~\eqref{eq:z1} (a consequence of particle number conservation) under the condition that $\delta z_1=\delta \gamma_\bullet^\rho=0$.

Clearly, a full diagrammatic analysis of $\chi_{nn}$ for the disordered electron liquid requires further steps. Details can be found, for example, in Ref.~\onlinecite{Castellani84}.

\subsubsection{The heat density-heat density correlation function}

In this section, we will discuss the structure of the heat density-heat density correlation function. The discussion will be organized in the same way as for the density-density correlation function.  

The correlation function can be split into static and dynamical parts. The static part is defined as $\chi_{kk}^{st}=\chi_{kk}(\bfq\rightarrow 0,\omega=0)$, and we write
\be
\chi_{kk}(\bfq,\omega)=\chi_{kk}^{st}+\chi_{kk}^{dyn}(\bfq,\omega).
\ee
The static and dynamical parts take the following structure
\be
\chi_{kk}^{st}&=&-c_{0}T\gamma_\bullet^z,\label{eq:chikkst}\\
\chi_{kk}^{dyn}(\bfq,\omega)&=&-c_0  T (\bar{\gamma}^z_\triangleleft)^2 [i\omega \tilde{\mathcal{D}}_\xi(\bfq,\omega)],\label{eq:chikkdyn}
\ee
where the propagator
\be
\tilde{\mathcal{D}}_\xi(\bfq,\omega)=\frac{\tilde{\xi}^2}{\tilde{D}\bfq^2-i\tilde{z}\omega}\label{eq:hatD}
\ee
depends on the constants $\tilde{z}$ and $\tilde{\xi}^2$ and $\tilde{D}$ which have to be found during the process of calculation. The relation between these quantities and $z$, $\xi^2$ and $D$ introduced for the density-density correlation function will be clarified later.
Further, $\bar{\gamma}^z_\triangleleft$ is a correction to the frequency vertex. The static part of the correlation function, as follows from Eq.~\eqref{eq:twolimits}, is determined by the specific heat $c$ of the electronic system, which is known to acquire quantum corrections within the renormalization group interval of energies. In Eq.~\eqref{eq:chikkst}, these corrections are absorbed into the quantity
\be
\gamma^z_\bullet=\frac{c}{c_0}.
\ee

\begin{figure}
\includegraphics[width=3.2cm]{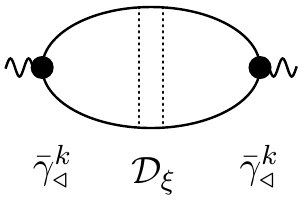}
\caption{The structure of the dynamical part of the heat density-heat density correlation function in accordance with Eq.~\eqref{eq:chikkdyn}.}
\label{fig:nncorrelationalt2}
\end{figure}

The structure of the dynamical part of the heat density-heat density correlation function is displayed in Fig.~\ref{fig:nncorrelationalt2}. Its \emph{main difference} from $\chi_{nn}^{dyn}$ is that for $\chi_{kk}^{dyn}$ ladder diagrams with the interaction amplitudes $\bar{\Gamma}_\rho$ as shown in Fig.~\ref{fig:nncorrelation} are not relevant. Therefore, the singularity of this correlation function is determined by the denominator of the diffuson propagator $\tilde{\mathcal{D}}_\xi$ without insertions describing rescattering; compare Fig.~\ref{fig:nncorrelationalt2} to Fig.~\ref{fig:nncorrelation}. The reason underlying this observation can be understood easily. An insertion of the static amplitude $\bar{\Gamma}_\rho$ decouples the frequency integrations on the left and right hand side of the diagram. As a consequence, a frequency integral of the type $\int_\eps \eps (\mathcal{F}_{\eps+\omega/2}-\mathcal{F}_{\eps-\omega/2})=0$ arises from the vertex related to the classical component of the gravitational potential, and diagrams with $\bar{\Gamma}_\rho$-insertion do not contribute to $\chi_{kk}^{dyn}$. It is important to note, however, that this simple observation does not imply the absence of vertical diagrams in general, as will be discussed in detail in the next section.

Energy conservation imposes a constraint on the correlation function $\chi_{kk}$, which is encoded in the relation $\chi_{kk}(\bfq=0,\omega\rightarrow 0)=0$, compare Eq.~\ref{eq:twolimit}. This, in turn, imposes the following constraint on the parameters entering the static and dynamical correlation functions
\be
\gamma_\bullet^z \tilde{z}=\tilde{\xi}^2(\bar{\gamma}_\triangleleft^z)^2.\label{eq:hatz}
\ee
As has been discussed in Refs.~\onlinecite{Castellani86}, the quantity $z$ which describes renormalization of the frequency term in action $S$, and in this way enters the propagator of diffusons, is directly related to the specific heat, $c=zc_0$, so that $\gamma_\bullet^z=z$. Using this information as an input, Eq.~\eqref{eq:hatz} can be also written as $z \tilde z=(\gamma_\triangleleft^z)^2$, where we defined $\gamma_\triangleleft^z=\tilde{\xi}\bar{\gamma}_\triangleleft^z$. Within the renormalization group interval of energies, this relation degenerates to $z=\tilde z=\gamma_\triangleleft^z$. Adding the static and the dynamical parts, one then finds
\be
\chi_{kk}(\bfq,\omega)=-\gamma_\bullet^z c_0T\frac{\tilde{D}\bfq^2}{\tilde{D}\bfq^2-i\tilde{z}\omega}.\label{eq:generalformchikk}
\ee

At the scale $1/\tau$, the initial values for the various parameters of the theory are $\gamma_\bullet^z=\bar{\gamma}_\triangleleft^z=\tilde{z}=\tilde{\xi}^2=1$, and the propagator of the diffuson is equal to
$\mathcal{D}(\textbf q, \omega)$, compare Eqs.~\eqref{eq:Dxi} and \eqref{eq:hatD} with Eq.~\eqref{eq:Dfree}. Coming back to the dynamical part, Eq.~\eqref{eq:chikkdyn}, we therefore expect that a perturbative calculation of the dynamical part of the correlation function will result in an expression of the following form
\be
\chi_{kk}^{dyn}(\bfq,\omega)&\approx&-c_0T\frac{i\omega(1+\delta \tilde{\xi}^2+2\delta\bar{\gamma}^z_\triangleleft)}{(D+\delta \tilde{D})\bfq^2-i(1+\delta \tilde{z})\omega}\no\\
&\approx&-c_0Ti\omega \mathcal{D}_{\textbf q, \omega}\no\\
&&-c_0Ti\omega (\delta \tilde{\xi}^2+2\delta \bar{\gamma}^z_{\triangleleft})\mathcal{D}_{\textbf q, \omega}\no\\
&&-c_0Ti\omega [-\delta \tilde{D}\bfq^2+i\omega\delta \tilde{z}]\mathcal{D}^2_{\textbf q, \omega}.\label{eq:kkmodel}
\ee
To check consistency of the sum of dynamical and static parts of $\chi_{kk}(\bfq,\omega)$ with the conservation laws, one should make certain, in view of Eq.~\eqref{eq:hatz}, that the relation $\delta \tilde{\xi}^2+2\delta \bar{\gamma}^z_{\triangleleft}-\delta \tilde z=\delta z$ indeed holds.

In the next section, the analysis of the logarithmic corrections to $\chi_{kk}$ as well as $\chi_{nn}$ is presented. In particular, in the following sections, Sec.~\ref{subsec:noninteracting} and Sec.~\ref{subsec:diagrams}, the structure of the different terms is discussed together with their diagrammatic representation, while in Sec.~\ref{subsec:Logarithmic} logarithmic corrections arising from the RG and sub-temperature intervals are described in detail. In Appendix~\ref{sec:listlogs} a comprehensive list of different contribution is given.

\section{Dynamical correlation functions - diagrammatic analysis and logarithmic corrections}
\label{sec:DynamicP}

In this Section, we present an analysis of the dynamical part of the heat density-heat density correlation function $\chi^{dyn}_{kk}$ in the diffusive limit. The analysis will be based on the NL$\sigma$M action derived in Sec.~\ref{sec:CoulombNLSM}, Eq.~\eqref{eq:actioneta}. To highlight similarities and differences, we contrast the calculation of $\chi_{kk}^{dyn}$ with that of $\chi^{dyn}_{nn}$ within the same formalism. In order to prepare the discussion of the interaction corrections, we first summarize the results for the non-interacting case.

\subsection{The non-interacting part of the correlation function}
\label{subsec:noninteracting}
\begin{figure}
\includegraphics[width=3.5cm]{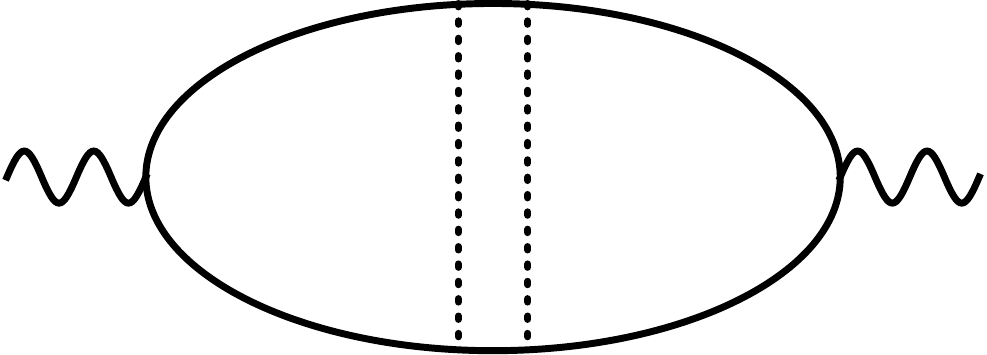}
\caption{Diagram for the non-interacting part of the dynamical correlation functions $\chi^{dyn}_{\eps\eps,0}$ and $\chi^{dyn}_{nn,0}$.}
\label{fig:chi0}
\end{figure}

In the absence of interactions, only the frequency-frequency correlation function $\chi^{dyn}_{\eps\eps}$ contributes to $\chi_{kk}^{dyn}$,
\begin{align}
&\chi^{dyn}_{\eps \eps,0}(x_1,x_2)=-\frac{i}{2}(\pi\nu_0)^2\int_{\eps_i}\mbox{e}^{-it_1(\eps_1-\eps_2)+it_2(\eps_3-\eps_4)}\no
\\&\times \langle\overline{\eps}_{12}\tr[\hat{\gamma}_2\underline{\sigma_3 \hat{P}_{\eps_1\eps_2}}(\bfr_1)]\overline{\eps}_{34}\tr[\hat{\gamma}_1\underline{\sigma_3 \hat{P}_{\eps_4\eps_3}}(\bfr_2)]\rangle_0.
\end{align}
The corresponding diagram is displayed in Fig.~\ref{fig:chi0}.
With the help of the contraction rules \eqref{eq:dst} or \eqref{eq:contr} on finds
\be
\chi^{dyn}_{\eps\eps,0}(\bfq,\omega)=-2i\pi\nu_0 \mathcal{D}_{\bfq,\omega}\int_\eps \eps^2\Delta_{\eps,\omega}.
\ee
Here, we introduced the window function
\be
\Delta_{\eps,\omega}=\mathcal{F}_{\eps+\omega/2}-\mathcal{F}_{\eps-\omega/2}.
\ee
The appearance of the window function is characteristic for the dynamical part of the correlation function. For $T\rightarrow 0$, it allows frequencies $\eps$ to lie in the interval $(\eps-\omega/2,\eps+\omega/2)$; at finite temperature this range broadens. Still, upon integration in $\eps$, the function $\Delta_{\eps,\omega}$ gives rise to the factor of $\omega$. Returning to the calculation of $\chi_{\eps\eps,0}^{dyn}$, after expansion in $\omega$ and with the help of the relation $\int_\eps \eps^2\partial_\eps \mathcal{F}_\eps=\pi T^2/3$, one obtains
\be
\chi^{dyn}_{\eps\eps,0}(\bfq,\omega)=-c_0 T i\omega\mathcal{D}_{\bfq,\omega},
\ee
where we remind that $c_0=2\pi^2\nu_0 T/3$ is the specific heat in the absence of quantum corrections.

In complete analogy, one can calculate the dynamical part of the density-density correlation function in the non-interacting limit,
\be
\chi_{nn,0}^{dyn}(\bfq,\omega)=-2i\pi\nu_0 \mathcal{D}_{\bfq,\omega}\int_\eps \Delta_{\eps,\omega}.
\ee
Using the relation $\pi \int_\eps \Delta_{\eps,\omega}=\omega$, one finds
\be
\chi_{nn,0}^{dyn}(\bfq,\omega)=-2\nu_0 i\omega\mathcal{D}_{\bfq,\omega}.
\ee
The diagrammatic representation for $\chi_{nn,0}^{dyn}$ coincides with the one for $\chi_{\eps\eps,0}^{dyn}$, compare Fig.~\ref{fig:chi0}. This is the origin of the WFL in the case of non-interacting electrons.

\subsection{Interaction corrections: Diagrams}
\label{subsec:diagrams}
We now turn to the explicit calculation of quantum corrections to the correlation functions originating from the combined effect of the long-ranged Coulomb interaction and disorder. The calculation is performed using an expansion in deviations $\delta Q$ from the saddle point, and applying subsequently the contraction rules formulated in Appendix~\ref{app:Contractions}. Diagrams are presented only for illustration. A detailed account of the calculation is presented in Appendix~\ref{sec:listlogs}. Here, we will highlight the most important diagrams and summarize the results. For comparison, we present the information for $\chi_{kk}$ in parallel with $\chi_{nn}$.

\begin{figure}
\includegraphics[width=7.5cm]{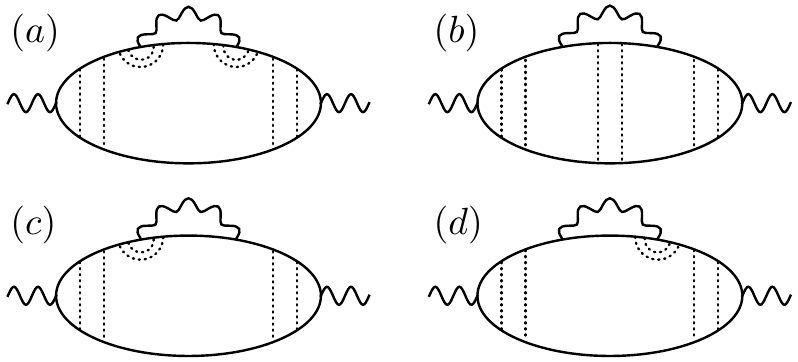}
\caption{The four horizontal diagrams contributing to $\chi^{dyn}_{kk,1}$ and $\chi^{dyn}_{nn,1}$. Each diagram has a symmetry-related partner that is not displayed here but accounted for in the analytical expressions discussed in the text.}
\label{fig:chi1}
\end{figure}

\begin{figure}
\includegraphics[width=3.5cm]{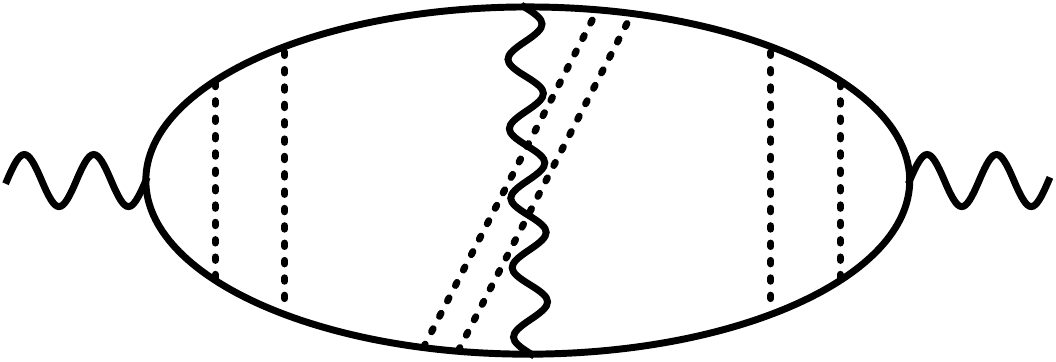}
\caption{The vertical diagram which contributes - together with its symmetry related partner - to $\chi^{dyn}_{kk,2}$ and $\chi_{nn,2}^{dyn}$.}
\label{fig:chi2}
\end{figure}

\begin{figure}
\includegraphics[width=7.5cm]{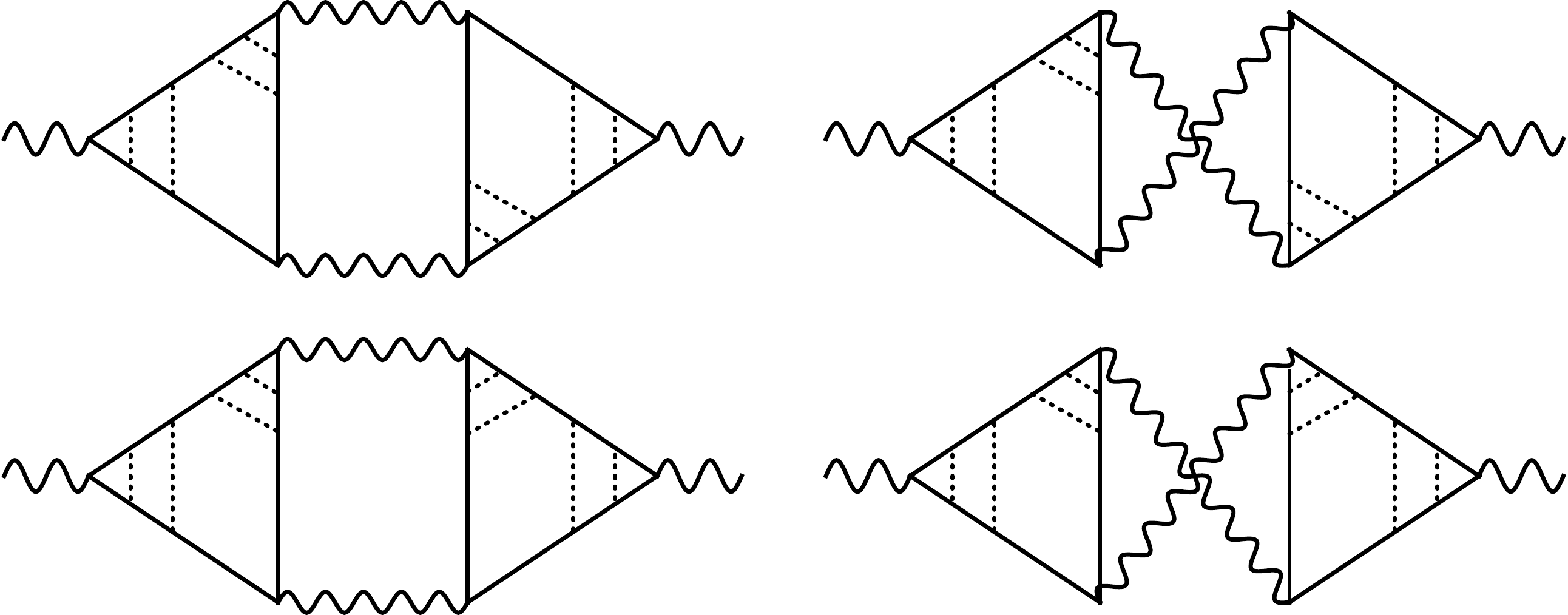}
\caption{The four drag diagrams which contribute to $\chi^{dyn}_{kk,3}$ and $\chi^{dyn}_{nn,3}$ together with their symmetry-related partners.}
\label{fig:chi3}
\end{figure}

\begin{figure}
\includegraphics[width=7.5cm]{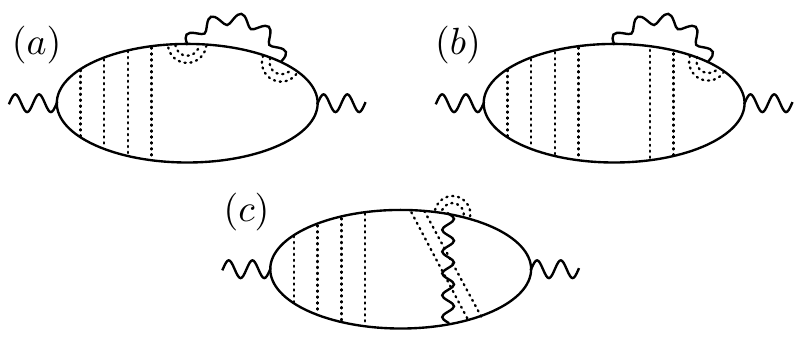}
\caption{The regular vertex corrections, $\chi^{dyn}_{kk,4}$ and $\chi^{dyn}_{nn,4}$. Three more diagrams are obtained by symmetrization.}
\label{fig:chi4}
\end{figure}

\begin{figure}
\includegraphics[width=7.5cm]{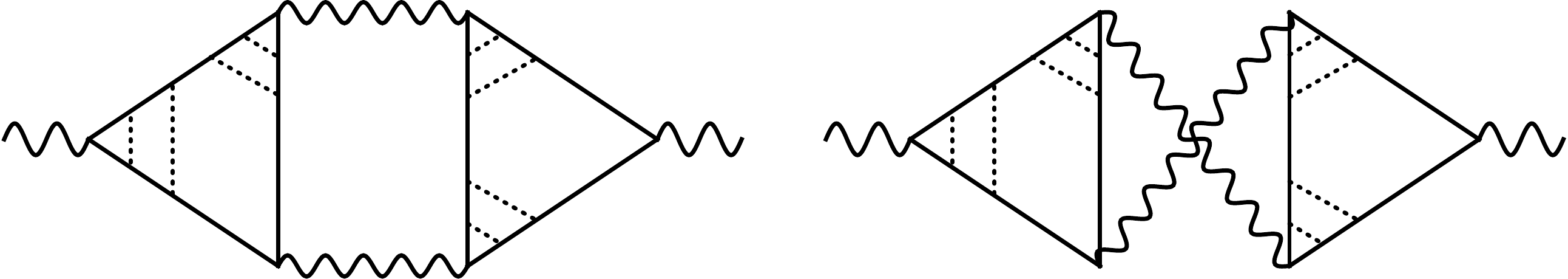}
\caption{Drag-type diagrams for the regular vertex corrections. An explicit calculation shows that their contribution to both $\chi_{kk,4}^{dyn}$ and $\chi_{nn,4}^{dyn}$ vanishes.}
\label{fig:Dragvertex}
\end{figure}

\begin{figure}
\includegraphics[width=7.5cm]{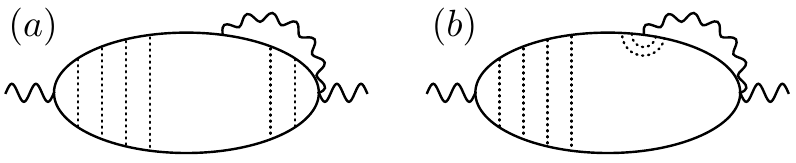}
\caption{Two anomalous vertex corrections contributing to $\chi_{kk,5}^{dyn}$. No analog exists for $\chi_{nn}^{dyn}$. Two more diagrams are obtained by symmetrization.}
\label{fig:chiV1}
\end{figure}

\begin{figure}
\includegraphics[width=7.5cm]{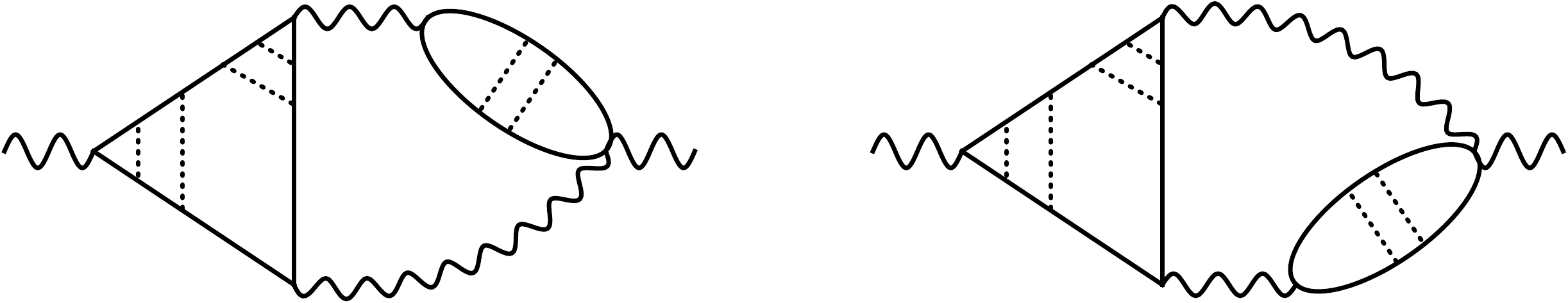}
\caption{Drag-type contributions to the anomalous vertex corrections $\chi_{kk,6}^{dyn}$. No analog exists for $\chi_{nn}^{dyn}$. Two more diagrams are obtained after symmetrization.}
\label{fig:chiV2}
\end{figure}

We will group the relevant diagrams for the calculation of the correlations functions into five classes. When we draw the diagrams, we leave out additional partner diagrams that can be obtained by simple symmetrization of those already displayed.
\begin{enumerate}
\item Horizontal diagrams: These diagrams contain a horizontal interaction line and give rise to corrections to the diffusion propagator. Vertex corrections with horizontal interaction lines will be considered separately. The horizontal diagrams are displayed in Fig.~\ref{fig:chi1}. The corresponding corrections will be labeled as $\chi_{kk,1}^{dyn}$ or $\chi_{nn,1}^{dyn}$.
\item Vertical diagram: The diagram with vertical interaction line relevant for our calculation is displayed in Fig.~\ref{fig:chi2}. It results in corrections to the diffusion propagator. Vertex corrections with vertical interaction lines will be considered separately. The vertical diagram leads to the corrections $\chi^{dyn}_{kk,2}$ and $\chi^{dyn}_{nn,2}$.
\item Drag diagrams: The drag diagrams contain two screened interaction lines and give rise to corrections to the diffusion propagator, see Fig.~\ref{fig:chi3}. The resulting corrections will be labeled as $\chi^{dyn}_{kk,3}$ and $\chi^{dyn}_{nn,3}$. Vertex corrections of drag-type will be considered separately.
\item Regular vertex corrections: In this class, we summarize those vertex corrections that originate from the frequency vertex $S_{\eta \eps}$. Horizontal and vertical (regular) vertex corrections are displayed in Fig.~\ref{fig:chi4}, (regular) vertex corrections of the drag type in Fig.~\ref{fig:Dragvertex}. The regular vertex corrections will be referred to as $\chi^{dyn}_{kk,4}$ and $\chi^{dyn}_{nn,4}$.
\item Anomalous vertex corrections: The anomalous vertex corrections result from the interaction vertex generated by $S_{\eta V}$. Obviously, they only arise in the calculation of the heat density-heat density correlation function. Fig.~\ref{fig:chiV1} shows the diagrams for anomalous vertex corrections with a single interaction line. They will be labeled as $\chi_{kk,5}^{dyn}$. Fig.~\ref{fig:chiV2} shows the diagrams for vertex corrections of the drag type, i.e., with two interaction lines. These corrections will be labeled as $\chi^{dyn}_{kk,6}$.
\end{enumerate}

\subsection{Analysis of logarithmic corrections}
\label{subsec:Logarithmic}

In this Section we will compare logarithmic corrections to the diffusion coefficient and the frequency renormalization arising in the heat density-heat density correlation function with those in the density-density correlation function. The logarithmic corrections can be classified according to the most important frequency and momentum regions in the integrals:

For \noindent(i)-a terms the frequency $\nu$ transferred by the electron interaction is larger than the electron frequency $|\eps|$ as well as temperature. The frequency integrations are controlled by the combination $\mathcal{F}_{\eps+\nu}-\mathcal{F}_{\eps-\nu}$. The transferred frequency and momentum cover the whole RG-interval. Electron-hole pairs excited via the interaction are virtual and contributions from the sub-thermal region are insignificant.

For \noindent(i)-b terms the frequency transfer is limited by the combination $\partial_\nu (\mathcal{F}_{\eps+\nu}-\mathcal{F}_{\eps-\nu})$. In this case, the frequency transfer is insignificant but the momentum integration covers the whole RG-interval. These terms describe the modification of the interaction amplitudes by disorder.

Finally, there appear new contributions, (ii) terms, which are determined by the combination $\mathcal{F}_{\eps+\nu}+\mathcal{F}_{\eps-\nu}$. In this case, the transferred frequency is limited either by temperature or by $|\eps|$. Furthermore, the electron interaction enters the integrals via its imaginary part, $\mbox{Im}V^R_{\bfk,\nu}$. This, together with the fact that the transferred frequency is limited either by $\eps$ or temperature, indicates that inelastic processes intervene. The momentum integration is determined by small momenta.

Logarithmic integrals appearing in (i)-a and (i)-b terms will be denoted as $I_i$-integrals; see Sec.~\ref{subsubsec:LogarithmicRG} and Appendix~\ref{sec:listlogs}. They are well known from the previous RG studies of the the disordered electron liquid. In contrast to electric transport, the contributions (ii) are specific for thermal transport; they are important in the case of the Coulomb interaction when $\mbox{Im}V^R_{\bfk,\nu}$ is singular. For a given frequency $\nu$, most important momenta fulfill the inequality $|\nu|/(D\kappa_s)<k<\sqrt{|\nu|/D}$. In this interval, one can approximate the dynamically screened interaction as
\be
\mbox{Im}V^R_{\bfk,\nu}\approx -\frac{1}{2\nu_0}\frac{\nu}{D\bfk^2}.
\ee
Eventually, the bare $1/D\bfk^2$ singularity gives rise to logarithmic corrections. These logarithmic integrals will be denoted as $I^h_i$, see  Sec.~\ref{subsubsec:LogarithmicIs} and Appendix~\ref{sec:listlogs}. The index $h$ emphasizes their importance for heat transport.

Note that the interval $|\nu|/(D\kappa_s)<k<\sqrt{|\nu|/D}$ is also responsible for the double-logarithmic dependence of the tunneling density of states as well as other spurious corrections that appear in intermediate stages of the RG procedure (compare the integral $I_1$ introduced below). In the case of the (ii)-type integrals, however, only a single logarithm arises, because allowed frequencies $\nu$ are small, of the order of the temperature, while a double-logarithmic dependence is obtained for an (i)-a type integral $I_1$, where the frequency can take large values.

In summary, we encounter two different types of contributions.
For the first type, which includes (i)-a and (i)-b terms, at least one of the two energies $|\nu|$ and $D\bfk^2$ lies in the RG interval $(T,1/\tau)$ giving rise to logarithmic integrals $I_i$. These corrections are well studied for the case of the density-density response function, both on a diagrammatic level and on the level of the field-theoretic NL$\sigma$M. Concerning the heat density-heat density correlation function, a diagrammatic study has been presented in Ref.~\onlinecite{Castellani87}, while the NL$\sigma$M of Ref.~\onlinecite{Schwiete14a,Schwiete14b} focused on the RG in the disordered Fermi liquid, i.e., in a disordered system with short-range Fermi liquid-type corrections. A common result of these studies was that the logarithmic corrections originating from the RG interval for the heat density-heat density correlation function lead to the sequence of equalities $z=\gamma_\bullet^z=\tilde z=\gamma_\triangleleft^z$.

The second type of logarithmic corrections, the (ii)-terms which originate from sub-thermal frequencies $\nu<T$, are at the center of our interest here. For these corrections, the imaginary part of the dynamically screened interaction is
relevant.

Besides these two types, there are terms that could, in principle, introduce a mass into the diffuson. They will be denoted as $J$-terms, for details see appendix \ref{sec:listlogs}. Unlike $I_i$ and $I_i^h$, which enter the calculation of the dynamical part of the heat-density heat-density correlation function together with the factors $Dq^2$ or $\omega$ only, these terms are finite (i.e., they do not vanish) in the limit $(\bfq,\omega)\rightarrow 0$. The $J$-terms arise as fragments of individual diagrams, but they have to cancel in the overall result for the correlation functions of conserved quantities. If they would persist, this would lead to a violation of the conservation laws. The cancellation of these terms is intimately related to the balance between in and out-terms in the collision integral integrated over frequencies (and for the case of the heat density correlation function also weighted with frequency).

\subsubsection{The finite $J$-terms}

We start with the $J$-terms. For the corrections arising in the density-density correlation function one gets:
\begin{align}
{\chi_{nn,1}^{dyn}}_J&=-2\nu_0 i\omega J_1(\bfq,\omega)\mathcal{D}^2_{\bf{q},\omega},\no\\
{\chi_{nn,2}^{dyn}}_J&=-2\nu_0 i\omega J_2(\bfq,\omega)\mathcal{D}^2_{\bf{q},\omega}.\label{eq:J1J2}
\end{align}
These formulas are obtained from the expressions given in appendix \ref{sec:listlogs} after performing the integration in the electronic frequencies $\eps$. Upon expansion in $D\bfq^2$ and $\omega$, $J_i(\bfq,\omega)=J_i^{0}+J_i^{D} D\bfq^2-J_i^{\omega}i\omega$, one notices that $J_1$ and $J_2$ both contain non-vanishing constant parts $J_i^{0}$ and singular expansion coefficients $J_i^{D}$ and $J_i^{\omega}$. It turns out, however, that there is a full cancellation between horizontal and vertical diagrams: $J_2(\bfq,\omega)=-J_1(\bfq,\omega)$. In particular, the cancellation between $J_1^{0}$ and $J_2^{0}=-J_1^0$ ensures that the density-density correlation function remains gapless. It is instructive to interpret the cancellation in the limit $(\bfq,\omega)\rightarrow 0$ in the language of kinetics (for a more detailed discussion see appendix \ref{app:cancel}). It can be seen that it is a direct result of condition $\int_{\eps,\bfr}\delta I_{coll}(\eps,x)=0$ for the linearized collision integral, which ensures the conservation of the particle number in a kinetic formulation of the problem.

Next, let us look at the corrections arising in the heat density-heat density correlation function:
\begin{align}
{\chi_{kk,1}^{dyn}}_J&=-c_0Ti\omega J_1(\bfq,\omega)\mathcal{D}^2_{\bf{q},\omega},\no\\
{\chi_{kk,2}^{dyn}}_J&=-c_0Ti\omega(J_2(\bfq,\omega)+\tilde{J}_2)\mathcal{D}^2_{\bf{q},\omega},\no\\
{\chi_{kk,3}^{dyn}}_J&=-c_0Ti\omega J_3\mathcal{D}^2_{\bf{q},\omega}.\label{eq:J1J3}
\end{align}
Here, the new terms $\tilde{J}_2$ and $J_3$, are finite and do not contain $\bfq$ and $\omega$ dependent parts. The cancellation of the $J$-terms, which results from the identities $\tilde{J}_2=-J_3$ in addition to $J_2=-J_1$, is now somewhat more complicated and involves the drag diagrams. In the present case, the cancellation is not guided by the number conservation or, equivalently, the absence of a mass of the diffuson, but by energy conservation. In the language of kinetics it can be seen that the cancellation is a direct consequence of the condition $\int_{\eps,\bfr}\eps\delta I_{coll}(\eps,x)=0$ for the linearized collision integral, which ensures the conservation of energy.

Due to the importance of the identity $J_1=-J_2$, and a similar relation $\tilde{J}_2=-J_3$ for the heat density-heat density correlation function, we devote Appendix \ref{app:cancel} to a more detailed discussion of this point. This discussion elucidates the relation between the horizontal, vertical and drag-type diagrams in the low-energy interval.

\subsubsection{Logarithmic corrections from the RG interval}
\label{subsubsec:LogarithmicRG}
Here, we list the logarithmic corrections to $\chi_{nn,0}^{dyn}=-2\nu i \omega \mathcal{D}$ originating from the RG interval. They are encoded in the logarithmic integrals denoted as $I_i$. Detailed derivations as well as the definitions of the appearing integrals $I_i$ can be found in Appendix \ref{sec:listlogs}
\begin{align}
\chi_{nn,1}^{dyn}&=-2\nu_0 i\omega\left[-2(D\bfq^2-i\omega)I_1+D\bfq^2 I_D-i\omega I_z\right]\mathcal{D}^2,\no\\
\chi_{nn,2}^{dyn}&=-2\nu_0 i\omega\left(i\omega I_z\right)\mathcal{D}^2,\no\\
\chi_{nn,3}^{dyn}&=0,\no\\
\chi_{nn,4}^{dyn}&=-2\nu_0 i\omega I_1\mathcal{D}.\label{eq:nnresults}
\end{align}
In this list, we suppressed the arguments of $\mathcal{D}$ for the sake of brevity. Concerning the vertex correction $\chi_{nn,4}^{dyn}$, the list presents the correction for one individual vertex, i.e., the sum of the vertex corrections from the left and the right vertex is twice as large.

For the heat density-heat density correlation function, we find the following corrections to $\chi_{kk,0}^{dyn}=-c_0Ti\omega\mathcal{D}_{\bfq,\omega}$:
\begin{align}
\chi_{kk,1}^{dyn}&=-c_0Ti\omega\left[-2(D\bfq^2-i\omega)I_1+D\bfq^2 I_D-i\omega I_z\right]\mathcal{D}^2,\no\\
\chi_{kk,2}^{dyn}&=-c_0Ti\omega\left[i\omega (I_z-I_2)\right]\mathcal{D}^2,\no\\
\chi_{kk,3}^{dyn}&=0,\no\\
\chi_{kk,4}^{dyn}&=-c_0T i\omega I_1\mathcal{D},\no\\
\chi_{kk,5}^{dyn}&=-c_0T i \omega (-I_5)\mathcal{D},\no\\
\chi_{kk,6}^{dyn}&=0.\label{eq:kkresults}
\end{align}
We would like to stress that those integrals in Eqs.~\eqref{eq:nnresults} and \eqref{eq:kkresults}, which are denoted by the same names, are not only equal but determined by the same expressions. Similar to the case of $\chi_{nn}^{dyn}$, the list cites vertex correction for one individual vertex only.

Due to the presence of the window function $\Delta_{\eps,\omega}\approx \omega\partial_\eps \mathcal{F}_\eps$, one may set the electron frequency $|\eps|\approx T$ in the expressions determining the integrals $I_i$. Then one finds
\begin{align}
I_1&=\frac{1}{6}\rho\log\frac{1}{T\tau}\log\frac{D\kappa_s^2}{T},\\
I_D&=\rho\log\frac{1}{T\tau},\\
I_z&=I_2=I_5=\frac{1}{2}I_D,
\end{align}
with $\rho=(4\pi^2 \nu_0 D)^{-1}$. Whereas the main contribution for the momentum integral in $I_1$ comes from the interval $|\nu|/(D\kappa_s)<k<\sqrt{|\nu|/D}$, for the rest of the terms relevant momenta are such that $D\bfk^2>|\nu|$. Relevant frequencies are large $|\nu|>T$ for $I_1$ and $I_D$, but $|\nu|\lesssim T$ for $I_z$, $I_2$, and $I_5$.

\emph{Density-density correlation function:} Following the logic of Sec.~\ref{sec:gstructure}, we arrange the obtained corrections to $\chi_{nn}$ into the general form consistent with that of a correlation function of a conserved quantity.
Then, from comparison with the results listed in Eq.~\eqref{eq:nnresults}, and the expression Eq.~\eqref{eq:nnmodel}
we can find the corrections to the various constants characterizing this correlation function. First of all, we observe that the wave function renormalizations given by the $I_1$-term in $\chi_{nn,1}^{dyn}$, and the vertex corrections $\delta \bar{\gamma}^\rho_{\triangleleft}$ given by $\chi_{nn,4}^{dyn}$, cancel out:
\begin{align}
\delta \xi^2&=-2I_1\\
\delta \bar{\gamma}^\rho_{\triangleleft}&=I_1\label{eq:Ione}.
\end{align}
This ensures the absence of doubly-logarithmic corrections in $\chi_{nn}$. Furthermore, the effect of the frequency
renormalization, $\delta z=-I_z$, given by the last term in $\chi_{nn,1}^{dyn}$ is cancelled by that of the renormalized screened Coulomb interaction, $\delta \bar{\Gamma}_\rho=-I_z$, given by $\chi_{nn,2}$. The only effective correction which remains after the cancelations is a correction to the diffusion coefficient in $\chi_{nn,1}^{dyn}$:
\be
\delta D=-D I_D.
\ee

Thus, we reproduced the following (known) results:

\noindent 1. The density-density correlation function in the presence of quantum corrections (albeit ignoring Fermi-liquid type corrections) reads
\be
\chi_{nn}(\bfq,\omega)=-2\nu_0\frac{D_n\bfq^2}{D_n\bfq^2-i\omega},
\ee
where the diffusion of charges is governed by the charge diffusion constant $D_n=D+\delta D$.

\noindent 2. Electric conductivity can be found from the relation
\be
\sigma=-e^2\lim_{\omega\rightarrow 0}\lim_{\bfq\rightarrow 0}\left[\frac{\omega}{\bfq^2}\mbox{Im}\chi_{nn}(\bfq,\omega)\right]=2\nu_0 e^2 D_n.\label{eq:sigmagen}
\ee

As a consequence of Eq.~\eqref{eq:sigmagen}, the correction to conductivity is
\be
\frac{\delta \sigma}{\sigma}=\frac{\delta D}{D}=-I_D=-\rho\log\frac{1}{T\tau}.
\ee
In this way, one recovers the well-known Altshuler-Aronov correction to conductivity from the formalism. This correction originates from the RG interval of energies.

\emph{Heat density-heat density correlation function:} A comparison of the results listed in Eq.~\eqref{eq:kkresults} and the general corrections stated in Eq.~\eqref{eq:kkmodel} leads us to the following relations for the corrections originating from the RG interval
\be
\delta \tilde{\xi}^2&=&\delta {\xi}^2=-2I_1,\no\\
\delta \tilde{D}&=&\delta D=-DI_D,\no\\
\delta \tilde{z}&=&\delta z=-I_z,\no\\
\delta \bar{\gamma}_\triangleleft^z&=&I_1-I_5.\label{eq:kkcomp}
\ee
As it has been discussed in Sec.~\ref{sec:gstructure}, for the consistency of $\chi_{kk}(\bf{q},\omega)$ with the energy conservation law, the condition $\delta \tilde{\xi}^2+2\delta \bar{\gamma}^z_{\triangleleft}-\delta \tilde z=\delta z$ is necessary. This condition is fulfilled provided that $I_z=I_5$. While $I_z$ and $I_5$ are a priori different integrals, they do coincide with logarithmic accuracy and the relation stated in Eq.~\eqref{eq:hatz} holds. We would like to stress that $I_5$ originates from the anomalous vertex correction, which only exists for $\chi_{kk}^{dyn}$ (and is absent for $\chi_{nn}^{dyn}$) and, therefore, the presence of the source term $S_{\eta V}$ as already mentioned is a very important ingredient of the theory.

Thermal conductivity can be found from the formula
\be
\kappa=-\frac{1}{T}\lim_{\omega\rightarrow 0}\lim_{\bfq\rightarrow 0}\left(\frac{\omega}{\bfq^2}\mbox{Im}\left[\chi_{kk}(\bfq,\omega)\right]\right)=\frac{z}{\tilde{z}}c_0 \tilde{D}.\label{eq:kappa1}
\ee
In the last equality we used the form of the correlation function stated in Eq.~\eqref{eq:generalformchikk} as well as the relation $c=c_0z$ introduced before. This implies the relation
\be
\frac{\kappa}{\sigma T}=\frac{zc_0\tilde{D}}{\tilde{z}2\nu_0 e^2D_n T}=\frac{z\tilde{D}}{\tilde{z}D_n}\mathcal{L}_0,\label{eq:WFL}
\ee
where $\mathcal{L}_0=\pi^2/3e^2$ is the so-called Lorentz number.

We thus arrive at the following conclusions concerning the Wiedemann-Franz law:

\noindent 1. If there were no additional corrections from the sub-temperature interval, then the set of equations listed in \eqref{eq:kkcomp} would immediately lead us to the conclusion that the WFL is fulfilled. Indeed, as one can see from the second and third relation in \eqref{eq:kkcomp}, the equalities $z=\tilde{z}$ and $\delta\tilde{D}=\delta D$ hold. Then the WFL remains true even in the presence of the quantum corrections originating from the RG interval.

\noindent
2. In order to obtain a violation of the WFL, the inequality $\tilde{z}D_n\ne z\tilde{D}$ is required to hold.

\subsubsection{Logarithmic corrections from the sub-temperature interval}
\label{subsubsec:LogarithmicIs}
For each diagram, only the corrections from the sub-temperature energy interval are listed below:

\begin{align}
\chi_{kk,1}^{dyn}&=0,\no\\
\chi_{kk,2}^{dyn}&=-c_0Ti\omega\left[-(D\bfq^2-i\omega)\tilde{I}_2^h\right]\mathcal{D}^2,\no\\
\chi_{kk,3}^{dyn}&=-c_0Ti\omega\left[I_3^hD\bfq^2-I_2^h i\omega\right]\mathcal{D}^2,\no\\
\chi_{kk,4}^{dyn}&=-c_0T i\omega \frac{1}{2} I_4^h\mathcal{D},\no\\
\chi_{kk,5}^{dyn}&=0\no\\
\chi_{kk,6}^{dyn}&=-c_0T i\omega (-I_6^h)\mathcal{D}.\label{eq:chisub}
\end{align}
Note with respect to $\chi_{kk,4}^{dyn}$ and $\chi_{kk,6}^{dyn}$, that the list cites vertex corrections for one individual vertex only. Concerning the integrals $I_i^h$, we notice that again we can set $|\eps|\sim T$ and then
\be
\tilde{I}_2^h=\rho\log\frac{D\kappa_s^2}{T}
\ee
and $\tilde{I}_2^h=I_2^h=I_4^h=2I_3^h=2I_6^h \equiv I^h$. In these integrals, relevant momenta are in the interval $|\nu|/(D\kappa_s)<k<\sqrt{|\nu|/D}$ and relevant frequencies are small $|\nu|\lesssim T$; see Sec.~\ref{subsec:Logarithmic} for a general description of the $I^h_i$-terms, and Appendix~\ref{sec:listlogs} for their detailed analysis.

Since in this manuscript, we study only first order logarithmic corrections to the correlation function, the accuracy of the calculation is not sufficient to make a definite statement about the structure of the correlation function as a whole. In particular, unlike for the RG corrections, the classification of the corrections in terms of $\delta \tilde{D}$, $\delta \tilde{z}$ and $\delta \bar{\gamma}_\triangleleft$ is not unambiguous. This remains so even if we assume for the wave function renormalization that it is unchanged, $\delta \tilde{\xi}=\delta \xi$, and take into consideration that the specific heat $c/c_0=z$ is not affected by the sub-temperature corrections. There still remains a degree of freedom for $\gamma^z_{\triangleleft}$ and $\tilde{z}$ within Eq.~\eqref{eq:hatz}. The final result, of course, will not depend on the choice of presentation of the correlation function $\chi_{kk}$.

Here, we fix the ambiguity following the origin of corrections in Eq.~\eqref{eq:chisub}. Then, the vertex corrections given by $\chi_{kk,4}^{dyn}$ and $\chi_{kk,6}^{dyn}$ cancel in total, $\delta \bar{\gamma}^z_{\triangleleft}=0$. Next, the frequency corrections to $\delta \tilde{z}$ originating from the frequency terms in $\chi_{kk,2}^{dyn}$ and $\chi_{kk,3}^{dyn}$ also cancel, so that $\delta \tilde{z}=0$.
Thus, this procedure leads us to the following set of sub-temperature corrections:
\be
\delta \gamma^z_{\triangleleft}=\delta \bar{\gamma}^z_{\triangleleft}&=&0,\no\\
\delta \tilde{z}&=&0,\no\\
\delta \tilde {D}^h&=&\frac{1}{2} I^h.\label{eq:subset}
\ee
In the procedure chosen here for fixing parameters, the structure of the correlation function (i.e., the vertex corrections and frequency renormalization) are controlled by the RG-interval, while the heat diffusion constant besides the corrections from the RG interval acquires a special contribution from the sub-temperature energy range, $\delta \tilde {D}^h$.

Only corrections from the sub-temperature regime are discussed here, therefore we can set $\delta z=0$ when checking the consistency with the energy conservation law for $\chi_{kk}(\bf q, \omega)$, which reduces to $2\delta \gamma^z_{\triangleleft}-\delta \tilde z=0$.

Finally, according to Eq.~\eqref{eq:kappa1}, the correction to thermal conductivity reads
\be
\delta \kappa =-\frac{T}{6}\log\frac{1}{T\tau}+\frac{T}{12} \log \frac{D\kappa_s^2}{T}\label{eq:kappadelta}.
\ee

\subsubsection{Violation of the Wiedemann-Franz law}
From the results collected in this section we can draw the following conclusions:
\begin{enumerate}
\item The full heat density-heat density correlation function can be written as
\be
\chi_{kk}(\bfq,\omega)=-cT\frac{D_k\bfq^2}{D_k\bfq^2-i\omega},\label{eq:canonical}
\ee
where $D_k=(D_n +\delta {\tilde D}^h)/z$ is the heat diffusion constant, and $c=zc_0$. The form presented in Eq.~\eqref{eq:canonical} is \emph{canonical} for a correlation function of a density of a consered quantity in the presence of disorder. 
\item Comparing corrections to the heat and electric conductivities
\be
\frac{\delta \kappa}{\kappa}=\frac{\delta \sigma}{\sigma}+\frac{1}{2}I^h,\label{eq:kapparatio}
\ee
one gets that the Lorenz ratio is enhanced
\be
\frac{1}{\mathcal{L}_0}\frac{\kappa}{\sigma T}=1+\frac{1}{2}I^h.
\ee
with $I^h=\rho\log (D\kappa_s^2/T)>0$.
\end{enumerate}
The positive sign of the correction indicates that for the disordered electron gas with long range Coulomb interaction heat transport is more effective than the WFL suggests.

\section{Conclusion}
We conducted an analysis of the heat-density heat-density correlation function in order to obtain quantum corrections to the thermal conductivity of the disordered electron gas. Our analysis focused on the role of the long-range Coulomb interaction in the diffusive limit and combined effects originating from different energy scales. RG-type corrections arise from the energy interval from the elastic scattering rate down to temperature ($1/\tau\gg T$) and do not violate the WFL. Additional corrections to the thermal conductivity originate from the sub-temperature energy range. These corrections do not have an analog for electric conductivity and therefore violate the WFL. The resulting thermal conductivity exceeds the prediction of the WFL. In this sense, heat transport is more effective than charge transport.
This result should be contrasted with the case of the disordered Fermi liquid, a model system with short range interactions. In the disordered Fermi liquid the WFL law is obeyed.

As for the origin of the WFL violating corrections we would like to stress that (i) the scattering processes in question involve on-shell energies $\sim T$ as well as the sub-temperature energy range and (ii) the relevant logarithmic integrals contain the imaginary part of the dynamically screened Coulomb interaction $\mbox{Im} V^R(\bfk,\nu)$. We therefore conclude that inelastic scattering is at the origin of the violation of the WFL. We checked that the correction to the heat diffusion coefficient caused by the long range Coulomb interaction is not modified by Fermi liquid interaction amplitudes. We thereby expect that the answer obtained for the correction to thermal conductivity $\delta \kappa$ presented in Eq.~\eqref{eq:kappadelta} is final.

We studied thermal conductivity in a situation where mechanical work (e.g., radiation of acoustic waves) can be neglected. If one additionally takes the conservation of particle number into consideration, this implies that heat transport is to a large extent governed by energy conservation. Special care has been taken regarding the definition of the energy density in the presence of the long-range Coulomb interaction. As the energy density depends on the electric field, the natural definition of a local energy conservation law requires a three-dimensional setting. For finding the three-dimensional energy density we used the field-theoretic construction of the energy-momentum tensor in combination with the principle of gauge invariance, which was used to lift the remaining ambiguity. These considerations naturally led to the Belinfante energy-momentum tensor.\cite{Belinfante40,Greiner96} Finally, in order to define an effective two-dimensional energy density, we employed a projection onto the plane.

It is instructive to contrast the thermal conductivity in the diffusive limit studied in this paper with known results in the clean electron gas with Coulomb interactions\cite{Lyakhov03} or in the ballistic limit.\cite{Catelani05} In the latter cases, inelastic scattering processes are responsible for a \emph{decrease} of the thermal conductivity. In contrast, the corrections in the diffusive limit lead to an \emph{increase} of the thermal conductivity. The positive sign of the correction indicates that the incoming scattering processes are dominant. Loosely speaking, in the diffusive case with long-range Coulomb interaction, electrons can use the energy $\sim T$ from a remote region to facilitate heat transfer.

\section*{Acknowledgments}
The authors thank K. Behnia, M. Brando, C.~Fr\"a\ss dorf, M. Feigel'man, I.~Gornyi, I.~Gruzberg, G.~Kotliar, T.~Kottos, B.~Shapiro, A.~Mirlin, E.~Mishchenko, J. Schmalian, J.~Sinova and C. Strunk for discussions. The authors gratefully acknowledge the support by the Alexander von Humboldt Foundation. The work in Russia was supported by the Russian
Science Foundation under the grant No. 14-42-00044. AF acknowledges support from US DOE award DE-SC0014154.

\appendix

\section{On the gauge invariant definition of the heat density}
\label{sec:belinfante}

The purpose of this appendix is to derive gauge-invariant expressions for the heat density and the heat current in the presence of the long-range Coulomb interaction. To this end, we follow the general procedure for the construction of the Belinfante tensor,\cite{Belinfante40,Greiner96} which is used for the energy-momentum tensor in electrodynamics. We start with the Lagrangian density (Schr\"odinger field coupled to electromagnetic field) $\mathcal{L}=\mathcal{L}_{S}+\mathcal{L}_{EM}$, where
\be
\mathcal{L}_S&=&\frac{i}{2}\left[\psi^*\partial_t\psi-\partial_t\psi^*\psi\right]\\
&&-\frac{1}{2m}(i\nabla-q{\bf A})\psi^*(-i\nabla-q{\bf A}) \psi-q\phi\psi^*\psi\no
\ee
is the Lagrangian of the non-relativistic Schr\"odinger field $\psi$ with charge $q$ and mass $m$ coupled to the electromagnetic field $A^\mu=(\phi,{\bf A})$, and
\be
\mathcal{L}_{EM}&=&-\frac{1}{16\pi}F^{\mu\nu}F_{\mu\nu}
\ee
is the Lagrangian of the free electromagnetic field.\cite{Jackson75} The potentials $\phi$ and ${\bf A}$ are related to the electric and magnetic fields
\be
{\bf E}=-\nabla \phi-\partial_t{\bf A}, \quad {\bf B}=\nabla\times  {\bf A}.\label{eq:EB}
\ee
Also, $F^{\mu\nu}=\partial^\mu A^\nu-\partial^\nu A^\mu$ is the field strength tensor. The relativistic notation is used for convenience only. The equations of motion obtained by a variation of the action $S=\int dx \mathcal{L}$ with respect to $\psi^*$, $\psi$, $A^0=\phi$ and ${\bf A}$ give the Schr\"odinger equation
\be
i\partial_t\psi=\frac{1}{2}(-i\nabla-q{\bf A})^2\psi+q\phi\psi,
\ee
and its conjugate, and the Maxwell equations $\nabla{\bf E}=4\pi \rho$ and $\nabla\times {\bf B}-\partial_t{\bf E}=4\pi{\bf j}$, respectively. Here, we defined the charge density $\rho=q\psi^*\psi$ and the current density
\be
{\bf j}=\frac{q}{2m}\left[\psi^*(-i\nabla-q{\bf A})\psi+((i\nabla -q{\bf A})\psi^*)\psi\right].
\ee
The remaining two Maxwell equations, $\nabla {\bf B}=0$ and $\nabla\times {\bf E}+\partial_t{\bf B}=0$, are fulfilled automatically through \eqref{eq:EB}.

The canonical energy momentum tensor $\Theta^{\mu\nu}$ is obtained using the invariance of the action with respect to the translation $x'^\mu=x^\mu+\varepsilon^\mu$,
\be
\Theta^{\mu\nu}&=&\frac{\partial\mathcal{L}}{\partial(\partial_\mu\psi)}\partial^\nu\psi+\frac{\partial\mathcal{L}}{\partial(\partial_\mu\psi^*)}\partial^\nu\psi^*\no\\
&&+\frac{\partial\mathcal{L}}{\partial(\partial_\mu A^\sigma)}\partial^\nu A^{\sigma}-g^{\mu\nu}\mathcal{L},
\ee
where $g^{\mu\nu}=\mbox{diag}(1,-1,-1,-1)$. We know that $\partial_\mu \Theta^{\mu\nu}=0$. It means that for each $\nu$ we get a local conservation law (continuity equation). The conservation law related to the energy density is given by $\partial_\mu\Theta^{\mu0}=0$. We, therefore, should calculate the energy density $\Theta^{00}$ and the $i$-th component of the energy current $\Theta^{i0}$. One finds
\begin{align}
\Theta^{00}&=u_{\psi}+\phi \rho-\frac{1}{8\pi}{\bf E}^2+\frac{1}{4\pi}\left[-{\bf E}\partial_t{\bf A}+\frac{1}{2}{\bf B}^2\right],
\end{align}
where we defined
\be
u_{\psi}=\frac{1}{2m}(i\nabla-q{\bf A})\psi^*(-i\nabla-q{\bf A})\psi.
\ee
With the help of the Maxwell equations one can rewrite this result alternatively as
\be
\Theta^{00}&=&u_{\psi}+\frac{1}{8\pi} \left({\bf E}^2+{\bf B}^2\right)+\frac{1}{4\pi}\nabla(\phi{\bf E}).
\ee

Next, we turn to the components of $\Theta^{i0}$, for which we find
\be
\Theta^{i0}={\bf j}_{\psi}^{\varepsilon,i}-\frac{1}{4\pi}{\bf E}^i\partial_t\phi+\frac{1}{4\pi}({\bf B}\times \partial_t{\bf A})^i,\no
\ee
where
\begin{align}
{\bf j}_{\psi}^\varepsilon=-\frac{i}{2m}\Big[\partial_t\psi^*(-i\nabla-q{\bf A})\psi-(i\nabla-q{\bf A})\psi^*\partial_t\psi\Big].
\end{align}
Again, with the help of the Maxwell equations, one can find the alternative representation
\be
\Theta^{i0}&=&({\bf j}_{\psi}^{\varepsilon}-\phi{\bf j})^i+\frac{1}{4\pi}({\bf E}\times{\bf B})^i\\
&&+\frac{1}{4\pi}\left[\nabla \times (\phi{\bf B})-\partial_t({\bf E}\phi)\right]^i.\no
\ee

As is well known, there is a problem with the canonical energy-momentum tensor; it is neither symmetric nor gauge invariant. This is already obvious from the terms ${\bf E}\partial_t{\bf A}$ and $({\bf B}\times \partial_t{\bf A})$ in the expressions for $\Theta^{00}$ and $\Theta^{0i}$, respectively. Since the densities and currents are not defined uniquely, one can add a four-divergence to the energy-momentum tensor as
\be
T_{\mu\nu}=\Theta_{\mu\nu}+\partial^\sigma\chi_{\sigma\mu\nu},
\ee
where $\chi$ fulfills the two requirements that $\chi_{\sigma\mu\nu}=-\chi_{\mu\sigma\nu}$ and $\chi_{0k\nu}$ falls off fast enough at infinite spatial distances so that a certain surface terms vanish. One can therefore use the Belinfante tensor $T$ instead of the canonical energy-momentum tensor $\Theta$ and write
\be
T^{\mu\nu}=\Theta^{\mu\nu}+\frac{1}{4\pi}\partial_\sigma(F^{\mu\sigma} A^\nu).
\ee
Since $F$ is antisymmetric, the relation $\partial_\mu T^{\mu\nu}=0$ follows immediately. Noting that $\partial_\sigma(F^{0\sigma} A^0)=-\nabla (\phi{\bf E})$ and $\partial_\sigma(F^{i\sigma}A^0)=\partial_t(\phi {\bf E})^i-(\nabla \times (\phi{\bf B}))^i$ one finds
\be
T^{00}&=&u_{\psi}+\frac{1}{8\pi}\left({\bf E}^2+{\bf B}^2\right)\label{eq:T00},\\
T^{i0}&=&({\bf j}_{\psi}^{\varepsilon}-\phi{\bf j})^i+\frac{1}{4\pi}({\bf E}\times{\bf B})^i\label{eq:Ti0}.
\ee
Note that $u_{\psi}$ is a gauge-invariant quantity, as a local phase change of $\psi$, $\psi^*$ can be absorbed by ${\bf A}$. We conclude that the energy density $T^{00}$ is gauge invariant. As to the current,
it can be easily checked that the combination ${\bf j}_0^{\varepsilon,i}-\phi{\bf j}^i$ is also gauge invariant, i.e., the transformation $\psi\rightarrow \exp(i\chi)\psi$ can be compensated by ${\bf A}\rightarrow {\bf A}+q^{-1}\nabla\chi$ and $\phi\rightarrow \phi-q^{-1}\partial_t\chi$, and so is $T^{i0}$.

Note that in the absence of external fields, and neglecting fluctuating magnetic fields, which is a relativistic effect, we find agreement between the components $\Theta^{\mu 0}$ of the canonical energy-momentum tensor and the \emph{intermediate} expressions ${\bf j}_?^\varepsilon$ and $u_?$ considered in Appendix B of Ref.~\onlinecite{Catelani05}, as well as between our expressions \eqref{eq:T00} and $\eqref{eq:Ti0}$ for the gauge-invariant energy density and current and the \emph{final} expressions obtained in Ref.~\onlinecite{Catelani05}.

Next, we specialize on the Coulomb gauge, the gauge used in the main text. It is convenient to decompose ${\bf A}={\bf A}^\parallel+{\bf A}^\perp$, with $\nabla {\bf A}^\perp=0$ and $\nabla\times {\bf A}^\parallel=0$. The Coulomb gauge $\nabla {\bf A}=0$ eliminates the longitudinal degrees of freedom ${\bf A}^\parallel=0$, so that ${\bf A}={\bf A}^\perp$. The electric field ${\bf E}=-\nabla A_0-\partial_t{\bf A}$, in contrast, has both a longitudinal and a transversal part, ${\bf E}^\parallel=-\nabla A_0$ and ${\bf E}^\perp=-\partial_t{\bf A}^\perp$. Then, $A_0$ is determined by the Poisson equation and
\be
{\bf E}^\parallel=-\frac{1}{4\pi}\nabla \int d{\bf r'}\frac{\rho({\bf r'},t)}{|{\bf r}-{\bf r'}|}.
\ee
In the non-relativistic limit we may neglect ${\bf A}^\perp$ (so that ${\bf E}^\perp\rightarrow 0$ and ${\bf B}\rightarrow 0$). In this case one obtains
\be
T^{00}&=& \frac{1}{2m}\nabla\psi^*\nabla\psi+\frac{1}{8\pi}[{\bf E}^\parallel]^2\label{eq:T00final}.
\ee
This is the expression for the energy density that will form the starting point for our considerations in the main text.

\section{Contractions rules}
\label{app:Contractions}

We state here the contraction rules for Gaussian averages with the action $S_0$ of Eq.~\eqref{eq:S0}.
To begin with, the matrix $\hat{P}$ can be represented as a matrix in Keldysh space as\cite{Kamenev11}
\be
\hat{P}_{\eps\eps'}(\bfr)=\left(\ba{cc} 0& d^{cl}_{\eps\eps'}(\bfr)\\d^{q}_{\eps\eps'}(\bfr)&0\ea\right),\label{eq:param}
\ee
where $d^{cl/q}$ are two hermitian matrices. The elementary contraction derived from $S_0$ reads
\be
&&\left\langle d^{cl}_{\alpha\beta;\eps_1\eps_2}(\bfq) d_{\gamma\delta;\eps_3\eps_4}^q(-\bfq_1)\right\rangle\no\\
&=&-\frac{2}{\pi\nu_0}\mathcal{D}(\bfq,\omega)\delta_{\bfq,\bfq_1}
\delta_{\eps_1,\eps_4}\delta_{\eps_2,\eps_3}\delta_{\alpha\delta}\delta_{\beta\gamma}\label{eq:dst},
\ee
where $\omega=\eps_1-\eps_2$, $\delta_{\bfq,\bfq_1}=(2\pi)^d\delta({\bf q}-{\bf q}_1)$, $\delta_{\eps_1,\eps_2}=2\pi\delta(\eps_1-\eps_2)$, and $\alpha,\beta,\gamma,\delta$ are spin indices. The diffuson $\mathcal{D}$ was defined in Eq.~\eqref{eq:diffuson}.
Starting from Eq.~\eqref{eq:dst}, one can formulate two convenient contraction rules for the matrices $\hat{P}$. The first rule can be used when the two matrices $\hat{P}$ appear under two different traces
\be
&&\left\langle \tr\left[\hat{A}\hat{P}_{\eps_1\eps_2}(\bfr_1)\right]\tr\left[\hat{B}\hat{P}_{\eps_3\eps_4}(\bfr_2)\right]\right\rangle\label{eq:contr}\\
&=&-\frac{2}{\pi\nu_0}\tr\left[ \hat{A}^\perp\hat{\Pi}_{\eps_1\eps_2}(\bfr_1-\bfr_2)\hat{B}^\perp\right]\delta_{\eps_1,\eps_4}\delta_{\eps_2,\eps_3}.\no
\ee
Here, we denoted $\hat{A}^\perp=\frac{1}{2}(\hat{A}-\hat{\sigma}_3\hat{A}\hat{\sigma}_3)$, and
\be
\hat{\Pi}_{\eps+\frac{\omega}{2}\eps-\frac{\omega}{2}}(\bfq)=\left(\ba{cc}\mathcal{D}_{\bfq,\omega}&0\\0&\overline{\mathcal{D}}_{\bfq,\omega}\ea\right)
\ee
contains both the advanced and the retarded diffusions $\mathcal{D}$ and $\overline{\mathcal{D}}$, respectively. The following second contraction rule is useful when the two matrices $\hat {P}$ stand under the same trace
\be
&&\left\langle \tr\left[AP_{\eps_1\eps_2}(\bfr_1)BP_{\eps_3\eps_4}(\bfr_2)\right]\right\rangle\\
&=&-\frac{1}{\pi\nu_0}\left(\tr[A\hat{\Pi}_{\eps_1\eps_2}(\bfr_1-\bfr_2)]\tr[B]\right.\no\\
&&\left.\qquad-\tr[A\hat{\sigma}_3 \hat{\Pi}_{\eps_1\eps_2}(\bfr_1-\bfr_2)]\tr[B\hat{\sigma}_3]  \right)\delta_{\eps_1,\eps_4}\delta_{\eps_2,\eps_3}.\no
\ee

\section{A list of contributions to $\chi_{nn}^{dyn}$ and $\chi_{kk}^{dyn}$}
\label{sec:listlogs}

In this Appendix, we provide details for the calculation of $\chi^{dyn}_{kk,i}$ and $\chi^{dyn}_{nn,i}$. These represent the contributions of different diagrams to the dynamical parts of the heat-density heat-density and density-density correlation functions, respectively. In particular, the Appendix contains the definitions of the logarithmic integrals $I_i$, $I^h_i$ and $J_i$. A classification of the different types of logarithmic integrals was discussed in Sec.~\ref{subsec:Logarithmic}.

\subsection{Horizontal diagrams}
\label{sec:horizontal}

Here, we consider contributions to the correlation functions originating from the expressions $\chi_{\eps\eps}^{dyn}$ and $\chi_{nn}^{dyn}$. We specialize on those terms, whose diagrammatic representation contains a horizontal interaction line. They are depicted in Fig.~\ref{fig:chi1}. A few remarks concerning these terms are in order here: diagram $(a)$ contains a Hikami box. The interaction part of $S_{\eta=0}$ enters in the form $\langle\Tr[\underline{{\phi}}{\sigma}_3 {P}]\Tr[\underline{{\phi}}{\sigma}_3{P}]\rangle_\phi$, diagram $(b)$ contains $\langle \Tr[\underline{{\phi}}{\sigma}_3 {P}^2]\Tr[\underline{{\phi}}{\sigma}_3 {P}^2]\rangle_\phi$, and diagram $(c)$ $\langle \Tr[\underline{{\phi}} {\sigma}_3 {P}]\Tr[\underline{{\phi}}{\sigma}_3 {P}^3]\rangle_\phi$. Here, $\underline{{\phi}}=u\circ \phi\circ{u}$ (for the sake of notational simplicity matrices in Keldysh spaces are denoted without the hat symbol here), and we used the notation
\be
\left\langle \phi^i(x)\phi^j(x')\right\rangle_\phi=\frac{i}{2}\hat{V}^{ij}(x-x').
\ee
Each diagram displayed in Fig.~\ref{fig:chi1} has a symmetry-related partner that is not displayed, but will be included in the expressions stated below. A common characteristic of the terms corresponding to the diagrams of Fig.~\ref{fig:chi1} is that they contain two diffusions $\mathcal{D}_{\bfq,\omega}^2$ and, in the case of $\chi^{dyn}_{\eps\eps}$, also the factor $\eps^2$. The latter fact is the main distinction from the vertical diagrams to be discussed below.

The result for the horizontal diagrams before expansion in $D\bfq^2$ and $\omega$ reads
\begin{align}
\left[\ba{cc}\chi_{kk,1}\\\chi_{nn,1}\ea\right]^{dyn}_{\bfq,\omega}=&2\pi\nu_0 \mathcal{D}_{\bfq,\omega}^2\int_{\eps}\left[\ba{cc}\eps^2\\1\ea\right]\Delta_{\eps,\omega}\sum_{i=a-d}\mathcal{X}_{1i}(\eps,\bfq,\omega),
\end{align}
where
\begin{align}
\mathcal{X}_{1a}=&-\frac{1}{6}\int_{\bfk,\nu}(\mathcal{F}_{\eps_1}-\mathcal{F}_{\eps_2}+\mathcal{F}_{\eps_2+\nu}-\mathcal{F}_{\eps_1-\nu})V^R_{\bfk,\nu},\no\\
&\times \mathcal{D}_{\bfk,\nu}^2[D(\bfk^2+\bfq^2)-i(\nu+\omega)]\\
\mathcal{X}_{1b}=&-\frac{1}{2}\int_{\bfk,\nu}\left[2V^K_{\bfk,\nu}+(\mathcal{F}_{\eps_1}-\mathcal{F}_{\eps_2})V^R_{\bfk,\nu}\right.,\no\\
&\left.+(\mathcal{F}_{\eps_1+\nu}-\mathcal{F}_{\eps_2-\nu})V^A_{\bfk,\nu}\right]\mathcal{D}_{\bfk+\bfq,\nu+\omega}\\
\mathcal{X}_{1c}+&\mathcal{X}_{1d}=\frac{2}{3}\int_{\bfk,\nu}(\mathcal{F}_{\eps_1}-\mathcal{F}_{\eps_2}+\mathcal{F}_{\eps_2+\nu}-\mathcal{F}_{\eps_1-\nu})\no\\
&\times V^R_{\bfk,\nu}\mathcal{D}_{\bfk,\nu}.
\end{align}
After expansion in $D\bfq^2$ and $\omega$ one arrives at the simplified expressions
\begin{align}
\left[\ba{cc}\chi_{kk,1}\\\chi_{nn,1}\ea\right]^{dyn}_{\bfq,\omega}&=-2\pi i\nu_0 \mathcal{D}_{\bfq,\omega}^2\int_{\eps}\left[\ba{cc}\eps^2\\1\ea\right]\Delta_{\eps,\omega}\mathcal{T}_1(\eps,\bfq,\omega),\label{eq:chi1}
\end{align}
where
\begin{align}
\mathcal{T}_1=&-2(D\bfq^2-i\omega)I_1(\eps)+D\bfq^2I_D(\eps)-i\omega I_z(\eps)\no\\
&+{J}_1(\bfq,\omega,\eps).\label{eq:T1}
\end{align}
The logarithmic integrals $I_i$ are defined as
\begin{align}
I_1(\eps)&=\frac{-i}{6}\int_{\bfk,\nu}\;(\mathcal{F}_{\eps+\nu}-\mathcal{F}_{\eps-\nu})\;\mathcal{D}_{\bfk,\nu}^2V^R_{\bfk,\nu},\label{eq:I1}\\
I_D(\eps)&=\frac{-2i}{d}\int_{\bfk,\nu}\;(\mathcal{F}_{\eps+\nu}-\mathcal{F}_{\eps-\nu})\;D\bfk^2\mathcal{D}_{\bfk,\nu}^3V^R_{\bfk,\nu},\\
I_z(\eps)&=\frac{1}{2}\int_{\bfk,\nu}\;(\partial_\eps \mathcal{F}_{\eps+\nu}+\partial_\eps \mathcal{F}_{\eps-\nu})\mathcal{D}_{\bfk,\nu}\;\mbox{Re}V^R_{\bfk,\nu}.
\end{align}
Momenta $k$ and frequencies $\nu$ in these integrals fulfill the conditions $D\bfk^2<1/\tau$ and $|\nu|<1/\tau$, i.e., they are confined to the diffusive regime. Due to the presence of $\Delta_{\eps,\omega}$ in Eq.~\eqref{eq:chi1} important values of $|\eps|$ under the integral are smaller or of the order of the temperature $T$. All three integrals $I_1, I_D, I_z$ are proportional to the dimensionless resistance $\rho=({4\pi^2 \nu_0 D})^{-1}$, the small parameter of the theory.

Next, we turn to the logarithmic divergencies arising from these integrals. Important momenta for the $\bfk$-integral in $I_1$ lie in the range $|\nu|/(D\kappa_s)<k<\sqrt{|\nu|/D}$. One obtains
\be
I_1(\eps)&\approx& \frac{\pi\rho}{12}\int_{\nu}\frac{\mathcal{F}_{\eps+\nu}-\mathcal{F}_{\eps-\nu}}{\nu}\log\frac{D\kappa_s^2}{\nu}\no\\
&\approx&\frac{\rho}{6}\log\frac{1}{\mbox{max}(|\eps|,T)\tau}\log\frac{D\kappa_s^2}{\mbox{max}(|\eps|,T)\tau}.
\ee
As will be discussed below, the double-logarithmic divergence arising from $I_1$ cancels from the density-density and the heat-density heat-density correlation function after taking all corrections into account. For the integrals $I_D$ and $I_z$, relevant momenta lie in the interval $|\nu|<D\bfk^2<{1}/{\tau}$, and one finds
\begin{align}
I_D(\eps)&\approx \frac{\pi\rho}{2}\int_{\nu}\frac{\mathcal{F}_{\eps+\nu}-\mathcal{F}_{\eps-\nu}}{\nu}\approx \rho\log\frac{1}{\mbox{max}(|\eps|,T)\tau},\\
I_z(\eps)&\approx \frac{\pi\rho}{2}\int_{\nu}\partial_\eps \mathcal{F}_{\eps+\nu}\log\frac{1}{|\nu|\tau}\approx\frac{\rho}{2}\log\frac{1}{\mbox{max}(|\eps|,T)\tau}.\no
\end{align}

In contrast to the three $I$-terms in Eq.~\eqref{eq:T1}, the term $J_1$ does not vanish in the limit $(\bfq,\omega)\rightarrow 0$. With the accuracy relevant for the present calculation, it can be written as
\begin{align}
&J_1(\bfq,\omega,\eps)=\no\\
&\int_{\bfk,\nu}\left[\mathcal{B}_\nu-\frac{1}{2}(\mathcal{F}_{\eps+\nu}-\mathcal{F}_{\eps-\nu})\right]\mathcal{D}_{\bfk+\bfq,\nu+\omega}\mbox{Im}V^R_{\bfk,\nu}.\label{eq:J1}
\end{align}
The subsequent integration in the electronic frequencies $\eps$ is controlled by the window function $\Delta_{\eps,\omega}\approx \omega\partial_\eps\mathcal{F}$. Therefore, in order to obtain $J_1(\bfq,\omega)$, the integral $J_1(\bfq,\omega,\eps)$ may be evaluated at $|\eps|\approx T$. The same remark applies to $J_2(\bfq,\omega)$ as well as the integrals $\tilde{J}_2$ and $J_3$. We will return to the discussion of the $J$-terms later in Appendix~\ref{app:cancel}.

\subsection{Vertical diagrams}
\label{sec:vertical}
Here, we discuss the contribution corresponding to the diagram displayed in Fig.~\ref{fig:chi2} and its symmetric partner. Technically, their origin is the same as for diagram $1(b)$, i.e., the use of $\langle \Tr[\underline{\phi}\sigma_3P^2]\Tr[\underline{\phi} \sigma_3 P^2]\rangle_\phi$ in the expressions for $\chi_{\eps\eps}^{dyn}$ and $\chi_{nn}^{dyn}$. The bare result reads
\begin{align}
\left[\ba{cc}\chi_{kk,2}\\\chi_{nn,2}\ea\right]^{dyn}_{\bfq,\omega}=&2\pi\nu_0 \mathcal{D}_{\bfq,\omega}^2\int_{\eps,\bfk,\nu}\left[\ba{cc}\eps(\eps+\nu)\\1\ea\right]\Delta_{\eps,\omega}\mathcal{D}_{\bfk+\bfq,\nu+\omega}\no\\
&\times[V^K_{\bfk,\nu}-\mathcal{F}_{\eps_2+\nu}V^R_{\bfk,\nu}+\mathcal{F}_{\eps_1+\nu} V^A_{\bfk,\nu}].
\end{align}
An expansion up to first order in $D\bfq^2$ and $\omega$ gives
\begin{align}
\left[\ba{cc}\chi_{kk,2}\\\chi_{nn,2}\ea\right]^{dyn}_{\bfq,\omega}=-2\pi i \nu_0 \mathcal{D}^2_{\bfq,\omega}\int_\eps \Delta_{\eps,\omega}\left[\ba{cc} \eps^2\mathcal{T}_2^{kk}(\eps,\bfq,\omega)\\\mathcal{T}_2^{nn}(\eps,\bfq,\omega)\ea\right].
\end{align}
The quantity $\mathcal{T}_2^{nn}$ associated with the density-density correlation function reads
\be
\mathcal{T}_2^{nn}&=&i\omega I_z(\eps)+J_2(\bfq,\omega,\eps),
\ee
where $J_2=-J_1$ and $J_1$ was defined in Eq.~\eqref{eq:J1}. The quantity $\mathcal{T}_2^{kk}$ associated with the heat density-heat density correlation function contains additional terms
\begin{align}
\mathcal{T}_2^{kk}=&\mathcal{T}_2^{nn}-(D\bfq^2-i\omega)\tilde{I}_2^h(\eps)+D\bfq^2 I_{reg}(\eps)-i\omega I_2(\eps)\no\\
&+\tilde{J}_2(\eps).\label{eq:T2kk}
\end{align}
Let us comment on the appearance of these additional terms. If we denote the frequency associated with the right frequency vertex as $\eps$, then the left frequency vertex carries the frequency $\eps\pm \nu$ due to the finite frequency transfer $\nu$ flowing through the interaction line. Correspondingly, we can decompose the expression for $\chi_{kk,2}^{dyn}$ into a part that contains the factor $\eps^2$ and a second one that contains the factor $\eps\nu$. The former part is accounted for by $\mathcal{T}_{2}^{nn}$, the latter part gives rise to the additional terms in Eq.~\eqref{eq:T2kk}. The
integrals $I_2$ and $\tilde{I}_2^h$ are defined as
\begin{align}
I_2(\eps)&=-\frac{1}{2\eps}\int_{\bfk,\nu}\;\nu\;\partial_\eps(\mathcal{F}_{\eps+\nu}-\mathcal{F}_{\eps-\nu})\mathcal{D}_{\bfk,\nu}\;\mbox{Re}V^R_{\bfk,\nu},\\
\tilde{I}_2^h(\eps)&=\frac{1}{\eps}\int_{\bfk,\nu}\;\nu\;(\mathcal{F}_{\eps+\nu}+\mathcal{F}_{\eps-\nu})\;\mathcal{D}^2_{\bfk,\nu}\;\mbox{Im} V^R_{\bfk,\nu}.
\end{align}
The integral $I_{reg}(\eps)$ is not logarithmic (regular) and just listed for completeness.

Let us discuss the integrals $I_2$ and $\tilde{I}_2^h$ one by one. Relevant momenta in the integral $I_2$ are confined to the range $|\nu|<D\bfk^2<{1}/{\tau}$ and one finds with logarithmic accuracy
\begin{align}
I_2(\eps)&\approx -\frac{\pi}{4\eps}\rho\int_{\nu}\nu\partial_\eps(\mathcal{F}_{\eps+\nu}
-\mathcal{F}_{\eps-\nu})\log\frac{1}{|\nu|\tau}\no\\&\approx \frac{\rho}{2}\log\frac{1}{\mbox{max}(|\eps|,T)\tau}.
\end{align}
The integral $\tilde{I}_2^h$ is the first in a series of (ii)-type integrals that will be considered. As was explained in Sec.~\ref{subsubsec:LogarithmicIs}, these integrals are mostly determined by momenta in the interval $|\nu|/(D\kappa_s)<k<\sqrt{|\nu|/D}$, while the integration over the frequency $\nu$ is severely limited by the combination $\mathcal{F}_{\eps+\nu}+\mathcal{F}_{\eps-\nu}$. As a result one gets
\begin{align}
\tilde{I}_2^h\approx\frac{\pi\rho}{2\eps}\int_\nu (\mathcal{F}_{\eps+\nu}+\mathcal{F}_{\eps-\nu})\log\frac{D\kappa_s^2}{|\nu|}\approx \rho\log\frac{D\kappa_s^2}{\mbox{max}(|\eps|,T)}.
\end{align}

Finally, let us state the integral
\be
\tilde{J}_2(\eps)=\frac{1}{\eps}\int_{\bfk,\nu}\nu\;(\mathcal{F}_{\eps+\nu}+\mathcal{F}_{\eps-\nu})\mathcal{D}_{\bfk,\nu}\;\mbox{Im} V^R_{\bfk,\nu}.
\ee
This integral $\tilde{J}_2$ will be discussed further in Appendix~\ref{app:cancel}. Note that this term, unlike $J_1$ and $J_2$, does not depend on $\bf{q}$ and $\omega$.

\subsection{Drag diagrams}
The two classes of diagrams discussed in Sec.~\ref{sec:horizontal} and Sec.~\ref{sec:vertical} contain a single (screened) interaction line. In this section we will discuss so-called drag diagrams, see Fig.~\ref{fig:chi3}, which form a subclass of those diagrams with two (screened) interaction lines. The diagrams are generated from the expressions for $\chi_{\eps\eps}^{dyn}$ and $\chi_{nn}^{dyn}$ given in Eqs.~\eqref{eq:chiepseps} and \eqref{eq:chinn}. It turns out that the drag diagrams do not contribute to the dynamical density-density correlation function,
\be
\chi_{nn,3}^{dyn}=0.
\ee
The full result for the drag contribution to the heat-density heat-density correlation function reads
\begin{align}
\chi_{kk,3}^{dyn}(\bfq,\omega)=&-2i\pi\nu_0^2\mathcal{D}_{\bfq,\omega}^2\int_{\eps}\eps\Delta_{\eps,\omega} \int_{\bfk,\nu} \nu (\nu-\omega)\no\\&\times(\mathcal{F}_{\eps_1-\nu}+\mathcal{F}_{\eps_2+\nu})V^R_{\bfk,\nu}V^A_{\bfk-\bfq,\nu-\omega}\no\\
&\times\mathcal{D}_{\bfk,\nu}(\mathcal{D}_{\bfk,\nu}+\bar{\mathcal{D}}_{\bfk-\bfq,\nu-\omega}).\label{eq:chi3kk}
\end{align}
The following two identities were used to obtain this result
\begin{align}
\pi\int_{\eps}\left[\ba{cc} \eps\\1\ea\right]\left[\mathcal{F}_{\eps_1}+\mathcal{F}_{\eps_2}-\mathcal{F}_{\eps_1-\nu}-\mathcal{F}_{\eps_2+\nu}\right]=\left[\ba{cc}\nu(\nu-\omega)\\0\ea\right].
\end{align}
Here, $\eps_{1,2}=\eps\pm\omega/2$. The fermion frequency $\eps$ in Eq.~\eqref{eq:chi3kk} is associated with the right loop of the drag diagram, while the integration in the above identities runs over the fermion frequency associated with the left loop.

Upon expansion in $D\bfq^2$ and $\omega$ one finds
\be
\chi^{dyn}_{kk,3}=-2\pi\nu_0 i \mathcal{D}^2_{\bfq,\omega}\int_\eps \Delta_{\eps,\omega}\eps^2\mathcal{T}^{\eps\eps}_3(\eps,\bfq,\omega),
\ee
where
\begin{align}
\mathcal{T}^{kk}_3(\eps,\bfq,\omega)=I_3^h D\bfq^2-I_2^h i\omega+J_3(\eps).\label{eq:Tkk3}
\end{align}
The integrals $I_2^h$ and $I_3^h$ are rather complicated expressions resulting from the expansion of Eq.~\eqref{eq:chi3kk} in $\omega$ and $D\bfq^2$ and we refrain from displaying them here. With logarithmic accuracy, one finds
\be
I_3^h=\frac{1}{2}\tilde{I}_2^h,\qquad I_2^h=\tilde{I}_2^h.
\ee
Relevant momenta in these integrals lie in the interval $|\nu|/(D\kappa_s)<k<\sqrt{|\nu|/D}$, and they originate from energies smaller than temperature.

Among other terms, there also appears a finite piece in the expression for $\mathcal{T}^{kk}_3$, Eq.~\eqref{eq:Tkk3},
\begin{align}
J_3(\eps)=&\frac{2\nu_0}{\eps}\int_{\bfk,\nu} \nu^2(\mathcal{F}_{\eps+\nu}+\mathcal{F}_{\eps-\nu})\no\\
&\times\mathcal{D}_{\bfk,\nu}^RV^R_{\bfk,\nu}V^A_{\bfk,\nu}\mbox{Re}\mathcal{D}_{\bfk,\nu},\label{eq:J3}
\end{align}
which will be discussed in Appendix~\ref{app:cancel} together with the related terms $J_1$, $J_2$, and $\tilde{J}_2$.

\subsection{Regular vertex corrections}
The terms considered in this section are obtained from the expressions for $\chi^{dyn}_{\eps\eps}$ [Eq.~\eqref{eq:chiepseps}] and $\chi_{nn}^{dyn}$ [Eq.~\ref{eq:chinn}] by taking into account nonlinear terms in the expansion of $\delta \hat{Q}$ in $\hat{P}$-modes at the vertices. The corresponding diagrams are displayed in Figs.~\ref{fig:chi4} and \ref{fig:Dragvertex}. As it turns out, the drag-type diagrams of Fig.~\ref{fig:Dragvertex} vanish both for the density-density correlation function and for the heat-density heat-density correlation function.

The expression corresponding to diagram $4a$, see Fig.~\ref{fig:chi4} reads
\begin{align}
\chi_{kk,4a}^{dyn}(\bfq,\omega)=&-\frac{2}{3}\pi\nu_0 \mathcal{D}_{\bfq,\omega}\int_{\eps} \eps^2\Delta_{\eps,\omega}\\
&\times \int_{\bfk,\nu}(\mathcal{F}_{\eps_1-\nu}-\mathcal{F}_{\eps_2+\nu}-\Delta_{\eps,\omega})\mathcal{D}_{\bfk,\nu}^2 V^R_{\bfk,\nu}.\no
\end{align}
Due to a cancellation between the term corresponding to the horizontal diagram (4b) and the vertical diagram (4c) it is convenient to state the sum:
\begin{align}
&\chi_{kk,4b}^{dyn}(\bfq,\omega)+\chi_{kk,4c}^{dyn}(\bfq,\omega)\\
=&\pi\nu_0 \mathcal{D}_{\bfq,\omega}\int_{\eps} \Delta_{\eps,\omega}\int_{\bfk,\nu}[\eps(\eps-\nu)\mathcal{F}_{\eps_1-\nu}-\eps(\eps+\nu)\mathcal{F}_{\eps_2+\nu}\no\\
&-\eps^2\Delta_{\eps,\omega}]\mathcal{D}_{\bfk,\nu}\mathcal{D}_{\bfk+\bfq,\nu+\omega} V^R_{\bfk,\nu}.\no
\end{align}
Correspondingly, for the density-density correlation function we get the somewhat simpler expressions
\begin{align}
\chi_{nn,4a}^{dyn}(\bfq,\omega)=&-\frac{2}{3}\pi\nu_0 \mathcal{D}_{\bfq,\omega}\int_{\eps} \Delta_{\eps,\omega}\\
&\times \int_{\bfk,\nu}(\mathcal{F}_{\eps_1-\nu}-\mathcal{F}_{\eps_2+\nu}-\Delta_{\eps,\omega})\mathcal{D}_{\bfk,\nu}^2 V^R_{\bfk,\nu},\no
\end{align}
\begin{align}
&\chi_{nn,4b}^{dyn}(\bfq,\omega)+\chi_{nn,4c}^{dyn}(\bfq,\omega)=\pi\nu_0 \mathcal{D}_{\bfq,\omega}\int_{\eps} \Delta_{\eps,\omega}\\
&\times \int_{\bfk,\nu}[\mathcal{F}_{\eps_1-\nu}-\mathcal{F}_{\eps_2+\nu}-\Delta_{\eps,\omega}] \mathcal{D}_{\bfk,\nu}\mathcal{D}_{\bfk+\bfq,\nu+\omega} V^R_{\bfk,\nu}.\no
\end{align}
Unlike for the contributions considered in the previous sections, no expansion in $D\bfq^2$ and $\omega$ is needed for the vertex corrections for both $\chi_{nn}^{dyn}$ and $\chi_{kk}^{dyn}$; we may safely put $\bfq\rightarrow 0$, $\omega\rightarrow 0$.

The result can be written in the following form
\begin{align}
\left[\ba{cc}\chi_{kk,4}\\\chi_{nn,4}\ea\right]^{dyn}_{\bfq,\omega}=-2\pi i \nu_0 \mathcal{D}_{\bfq,\omega}\int_\eps \Delta_{\eps,\omega}\left[\ba{cc} \eps^2\mathcal{T}_4^{kk}(\eps,\bfq,\omega)\\\mathcal{T}_4^{nn}(\eps,\bfq,\omega)\ea\right],
\end{align}
where
\be
\mathcal{T}_4^{nn}&=&I_1(\eps),\\
\mathcal{T}_4^{kk}&=&I_1(\eps)+\frac{1}{2}I_4^h(\eps).
\ee
The logarithmic integral $I_1(\eps)$ was defined in Eq.~\eqref{eq:I1} and the new integral is
\be
I_4^h(\eps)=-\frac{i}{\eps}\int_{\bfk,\nu}\nu(\mathcal{F}_{\eps+\nu}+\mathcal{F}_{\eps-\nu})\mathcal{D}^2_{\bfk,\nu} V^R_{\bfk,\nu}.
\ee
Note that the same corrections also originate from the corresponding diagrams for the left vertex. Only $\mbox{Im}V^R_{\bfk,\nu}$ is relevant and one finds the same integral as for the vertical diagrams: $I_4^h=\tilde{I}_2^h$.

\subsection{Anomalous vertex corrections}
We refer to those vertex corrections that originate from $\chi_{\eps V}^{dyn}$ and $\chi_{V \eps}^{dyn}$ as anomalous. For an illustration, see Fig.~\ref{fig:chiV1}; no anomalous vertex corrections exist for the density-density correlation function. The analytical expressions are
\begin{align}
\chi^{dyn}_{kk,5a}(\bfq,\omega)=&\frac{i}{2}\pi\nu_0 \mathcal{D}_{\bfq,\omega}\int_{\eps}\eps\Delta_{\eps,\omega}\int_{\bfk,\nu}(\mathcal{F}_{\eps_1-\nu}+\mathcal{F}_{\eps_2+\nu})\no\\
&\times \overline{\mathcal{D}}_{\bfk-\bfq,\nu-\omega}V^R_{\bfk,\nu}\\
\chi^{dyn}_{kk,5b}(\bfq,\omega)=&\frac{i}{2}\pi\nu\mathcal{D}_{\bfq,\omega}\int_{\eps}\eps\Delta_{\eps,\omega}\int_{\bfk,\omega}(\mathcal{F}_{\eps_1-\nu}+\mathcal{F}_{\eps_2+\nu})\no\\
&\times \mathcal{D}_{\bfk,\nu}V^R_{\bfk,\nu}.
\end{align}
As we are dealing with vertex corrections, we may safely set $(\bfq,\omega)\rightarrow 0$. When combining these two results, one finds
\be
\chi^{dyn}_{kk,5}(\bfq,\omega)=-2\pi\nu_0 i\mathcal{D}_{\bfq,\omega}\int_\eps \Delta_{\eps,\omega}\eps^2 \mathcal{T}_5^{kk}(\eps)
\ee
with
\be
\mathcal{T}_5^{kk}=-I_5(\eps).
\ee
The logarithmic integral $I_5$ is defined as
\be
I_5(\eps)=\frac{1}{2\eps}\int_{\bfk,\nu}(\mathcal{F}_{\eps+\nu}+\mathcal{F}_{\eps-\nu})\mathcal{D}_{\bfk,\nu}\mbox{Re} V^R_{\bfk,\nu}.
\ee
The main contribution originates from large momenta $D\bfk^2>\nu$ and one easily finds
\be
I_5(\eps)=\frac{\rho}{2}\log\frac{1}{\max\{|\eps|,T\}\tau}.
\ee

Another anomalous vertex correction arises from the drag-type diagrams of Fig.~\ref{fig:chiV2} (note that only the correction to the $\gamma_1$ vertex is written here)
\begin{align}
\chi^{dyn}_{kk,6}(\bfq,\omega)=&\pi\nu_0^2\mathcal{D}_{\bfq,\omega}\int_\eps \Delta_{\eps,\omega}\int_{\bfk,\nu}\eps\nu(\mathcal{F}_{\eps_1-\nu}+\mathcal{F}_{\eps_2+\nu})\no\\
&\times V^R_{\bfk,\nu}V^A_{\bfk-\bfq,\nu-\omega}\mathcal{D}_{\bfk,\nu}(\mathcal{D}_{\bfk,\nu}+\overline{\mathcal{D}}_{\bfk-\bfq,\nu-\omega})
\end{align}
Putting $(\bfq,\omega)\rightarrow 0$, we find
\be
\chi^{dyn}_{kk,6}(\bfq,\omega)=-2\pi\nu_0 i\mathcal{D}_{\bfq,\omega}\int_\eps \Delta_{\eps,\omega}\eps^2 \mathcal{T}_6^{kk}(\eps),
\ee
where
\begin{align}
\mathcal{T}_6^{kk}&=-I_6^h(\eps),\\
I_6^h(\eps)&=-\frac{i\nu_0}{\eps}\int_{\bfk,\nu}\nu (\mathcal{F}_{\eps+\nu}+\mathcal{F}_{\eps-\nu})V^R_{\bfk,\nu}V^A_{\bfk,\nu}\mathcal{D}_{\bfk,\nu}\mbox{Re}\mathcal{D}_{\bfk,\nu}.
\end{align}
Using the relation
\be
-2\nu_0\nu V^R_{\bfk,\nu}V^A_{\bfk,\nu}\mbox{Re}\mathcal{D}_{\bfk,\nu}=\mbox{Im} V^R_{\bfk,\nu}\label{eq:auxrel}
\ee
one can transform the integral to
\be
I_6^h(\eps)=\frac{i}{2\eps}\int_{\bfk,\nu} (\mathcal{F}_{\eps+\nu}+\mathcal{F}_{\eps-\nu})\mathcal{D}_{\bfk,\nu}\mbox{Im} V_{\bfk,\nu}^R.
\ee
Since relevant momenta are in the interval $|\nu|/(D\kappa_s)<k<\sqrt{|\nu|/D}$, one finds with logarithmic accuracy $I_6^h=\frac{1}{2}\tilde{I}_2^h$.

Unlike $I_i$, which are determined by very different integrals, see Sec.~\ref{subsec:Logarithmic} for the general classification of the logarithmic integrals, all the integrals $I^h_i$ reduce to the same expression.

\section{Cancellation of finite $J$-terms and the collision integral}
\label{app:cancel}

We reinterpret in the language of kinetics the cancellation of the $J$-terms between horizontal and vertical diagrams for the density-density correlation function, and between vertical, horizontal and drag diagrams for the heat-density heat-density correlation function.

\subsection{The case of the density-density correlation function}

Consider the deviation of the density from its equilibrium value caused by a weak, slowly varying external potential. In linear response, this deviation is characterized by the density-density correlation function. Our goal is to compare the $J$-terms arising during the calculation of the density-density correlation function to the Coulomb collision integral, which is well known and reads \cite{Altshuler85,Kamenev11}
\begin{align}
&I_{coll}(\eps,x)=-2\int_{\bfk,\nu}\mbox{Re}\mathcal{D}_{\bfk,\nu}\mbox{Im}V^R_{\bfk,\nu}\\
&\times \left[1-\mathcal{F}_{\eps-\nu}(x)\mathcal{F}_\eps(x)-\mathcal{B}_\nu(x)(\mathcal{F}_{\eps}(x)-\mathcal{F}_{\eps-\nu}(x))\right],\no
\end{align}
where
\be
\mathcal{B}_\nu(x)=\frac{\pi}{\nu}\int_{\eps'}\left(1-\mathcal{F}_{\eps'}(x)\mathcal{F}_{\eps'-\nu}(x)\right).
\ee
This expression for $\mathcal{B}_\nu(x)$ is a generalization of the well known relation connecting the bosonic and fermionic equilibrium distribution functions, $\mathcal{B}_\nu=\frac{\pi}{\nu}\int_{\eps}(1-\mathcal{F}_{\eps}\mathcal{F}_{\eps-\nu})$.

In equilibrium, i.e., for $\mathcal{F}_{\eps}(x)\rightarrow\mathcal{F}_\eps$, the collision integral vanishes identically. Writing $\mathcal{F}_\eps(x)=\mathcal{F}_\eps+\delta \mathcal{F}_\eps(x)$, the linearized collision integral reads
\begin{align}
&\delta I_{coll}(\eps,x)=-2\int_{\bfk,\nu}\mbox{Re}\mathcal{D}_{\bfk,\nu}\mbox{Im}V^R_{\bfk,\nu}\left[\delta \mathcal{F}_{\eps-\nu}(x)(\mathcal{B}_\nu-\mathcal{F}_{\eps})\right.\no\\
&\left.-\delta \mathcal{F}_{\eps}(x)(\mathcal{B}_\nu+\mathcal{F}_{\eps-\nu})+\delta \mathcal{B}_\nu(x)(\mathcal{F}_{\eps-\nu}-\mathcal{F}_\eps)\right],\label{eq:linColl}
\end{align}
where
\be
\delta \mathcal{B}_\nu(x)=-\frac{\pi}{\nu}\int_{\eps'}\delta \mathcal{F}_{\eps'}(x)(\mathcal{F}_{\eps'+\nu}+\mathcal{F}_{\eps'-\nu}).\label{eq:deltaB}
\ee
In the language of kinetics the conservation of the number of particles requires the vanishing of $\int_{\bfr,\eps}\delta I_{coll}(\eps,x)$. We will explain here that the cancellation of $J$-terms originating from horizontal and vertical diagrams is a result of the relation $\int_{\bfr,\eps}\delta I_{coll}(\eps,x)=0$.

In an iterative approach to the kinetic problem, which corresponds to our perturbative treatment of the screened Coulomb interaction, we next use the change in the distribution function calculated \emph{in the absence} of interactions, $\delta \mathcal{F}^{(0)}_{\eps}$, as a zeroth order solution. It is easy to see that then $\delta\mathcal{F}^{(0)}_\eps$ is proportional to the window function $\Delta_{\eps,\omega}$ and that, therefore, the bosonic distribution function remains unchanged, $\delta \mathcal{B}_{\nu}^{(0)}(x)=0$.

It is now possible to establish a connection of $\int_{\eps}\delta I_{coll}$ with the diagrammatic calculation.
The term proportional to $\delta \mathcal{F}^{(0)}_{\eps}$ in the linearized collision integral \eqref{eq:linColl} evaluated for $\delta \mathcal{F}=\delta \mathcal{F}^{(0)}$ is related to the horizontal diagrams considered in Sec.~\ref{sec:horizontal}. In a similar vein, the term proportional to $\delta \mathcal{F}^{(0)}_{\eps-\nu}$ is related to the vertical diagrams, see Sec.~\ref{sec:vertical}. Finally, the vanishing of $\delta \mathcal{B}_{\nu}^{(0)}(x)$ is directly related to the absence of drag-type corrections for $\chi^{dyn}_{nn}$.

Turning more specifically to the question of number conservation, we next focus on the expression for $\int_\eps\delta I_{coll}^{(0)}$. For the purpose of comparison, we reproduce here the integral $J_1$, \eqref{eq:J1}, in the limit $(\bfq,\omega)\rightarrow 0$, which arises from the horizontal diagrams considered in Sec.~\ref{sec:horizontal}:
\be
J_1=\int_{\bfk,\nu}\left[\mathcal{B}_\nu-\frac{1}{2}(\mathcal{F}_{\eps+\nu}-\mathcal{F}_{\eps-\nu})\right]\mbox{Re}\mathcal{D}_{\bfk,\nu}\mbox{Im}V^R_{\bfk,\nu}.\label{eq:J1limit}
\ee
Now note that the term containing $\delta \mathcal{F}_\eps(x)$ in Eq.~\eqref{eq:linColl} is proportional to $J_1$. In order to see this clearly, one should symmetrize the expression in $\nu$, using the oddness of $\mbox{Re}\mathcal{D}_{\bfk,\nu}\mbox{Im}V^R_{\bfk,\nu}$. A similar operation has to be performed for the term generated by $\delta \mathcal{F}^{(0)}_{\eps-\nu}$ in order to see that it is related to $J_2=-J_1$. This operation consists of a frequency shift $\eps\rightarrow \eps+\nu$ \emph{under the integral} $\int_{\eps}$ and subsequent symmetrization in $\nu$. In summary, one finds that horizontal and vertical terms in the collision integral are determined by $J_1$ of Eq.~\eqref{eq:J1limit} and $J_2=-J_1$, respectively. The cancellation observed in the diagrammatic calculation therefore results from particle number conservation expressed though $\int_{\eps}I^{(0)}_{coll}(\eps,\bfq=0,\omega)=0$. In the context of the diagrammatic calculation, it manifests itself in the absence of a mass of the diffusion.

As a final remark on this topic let us note that the separation into horizontal and vertical diagrams does not correspond to the separation into out- and in-terms in the studied collision integral.

\subsection{The case of the heat density-heat density correlation function}

Let $\delta \mathcal{F}_{\eps}^{(0)}$ now be the perturbation caused by a smoothly varying gravitational potential calculated in the absence of interactions. In this case one finds $\delta \mathcal{F}^{(0)}_\eps\propto \eps \Delta_{\eps,\omega}$. As before, in $\delta I_{coll}^{(0)}$ terms with $\delta \mathcal{F}^{(0)}_{\eps-\nu}$ are related to the vertical diagrams, and terms with $\delta \mathcal{F}^{(0)}_{\eps}$ to the horizontal diagrams. Unlike for the density-density correlation function, however, $\delta \mathcal{B}_{\nu}^{(0)}$, which is related to the drag diagrams, does not vanish.

As we will explain in the remainder of this section, the cancellation of constant terms between horizontal, vertical and drag diagrams for the heat density-heat density correlation function is a result of the relation $\int_{\bfr,\eps}\eps \delta I_{coll}(\eps,x)=0$. As a first step, one obtains the following relation with the help of Eq.~\eqref{eq:linColl} and after shifting $\eps\rightarrow \eps+\nu$ in the expression containing $\delta \mathcal{F}^{(0)}_{\eps-\nu}$:
\begin{align}
&\int_\eps\eps \delta I_{coll}^{(0)}=-2\int_{\bfk,\nu}\mbox{Re}\mathcal{D}_{\bfk,\nu} \mbox{Im} V^R_{\bfk,\nu}\no\\
&\times \int_\eps[(\eps+\nu)\delta\mathcal{F}^{(0)}_{\eps}(\mathcal{B}_\nu-\mathcal{F}_{\eps+\nu})-\eps \delta \mathcal{F}^{(0)}_\eps(\mathcal{B}_\nu+\mathcal{F}_{\eps-\nu})\no\\
&\qquad+\eps \delta \mathcal{B}^{(0)}_\nu (\mathcal{F}_{\eps-\nu}-\mathcal{F}_\eps)].
\end{align}
The $x$-dependence of $\delta \mathcal{F}^{(0)}$ and $\delta \mathcal{B}^{(0)}$ was suppressed for the sake of brevity.

We see that the terms proportional to $\eps$ cancel between the first and the second term upon symmetrization in $\nu$. This is the cancellation between the horizontal and the vertical diagrams encountered before for the density-density correlation function. The $\mathcal{B}_\nu$ term cancels because it is antisymmetric in $\nu$. Further, we can use the identity $\int_\eps \eps (\mathcal{F}_{\eps-\nu}-\mathcal{F}_\eps)=-\nu^2/2\pi$ for the last term. After symmetrizing the remaining integrand with respect to $\nu\leftrightarrow -\nu$, one obtains
\begin{align}
&\int_\eps\eps \delta I_{coll}^{(0)}=2\int_{\bfk,\nu}\mbox{Re}\mathcal{D}_{\bfk,\nu} \mbox{Im} V^R_{\bfk,\nu}\label{eq:collaux}\\
&\qquad\times \left[\int_\eps \nu \delta\mathcal{F}^{(0)}_{\eps}\frac{1}{2}(\mathcal{F}_{\eps+\nu}+\mathcal{F}_{\eps-\nu})+ \frac{1}{2\pi}\nu^2\delta \mathcal{B}^{(0)}_\nu \right]=0.\no
\end{align}
The second equality in \eqref{eq:collaux} becomes obvious upon substituting the expression for $\delta\mathcal{B}^{(0)}_\nu$, compare Eq.~\eqref{eq:deltaB}. The first of the two terms in the integral displayed in Eq.~\eqref{eq:collaux} corresponds to $\tilde{J}_2$, i.e., it originates from the vertical diagrams, the second term corresponds to the contribution from the drag diagram, i.e, to $J_3$. To see this clearly, one should insert the identity \eqref{eq:auxrel} into the definition of $J_3$, Eq.~\eqref{eq:J3}, which becomes
\be
J_3=-\frac{1}{\eps}\int_{\bfk,\nu} \nu(\mathcal{F}_{\eps+\nu}+\mathcal{F}_{\eps-\nu})\mathcal{D}_{\bfk,\nu}\mbox{Im} V^R_{\bfk,\nu}=-\tilde{J}_2.
\ee
This concludes our discussion of the cancellation of constant terms for the calculation of the heat density-heat density correlation function.

%\bibliography{library-1,library-2}

\begin{thebibliography}{37}
\expandafter\ifx\csname natexlab\endcsname\relax\def\natexlab#1{#1}\fi
\expandafter\ifx\csname bibnamefont\endcsname\relax
  \def\bibnamefont#1{#1}\fi
\expandafter\ifx\csname bibfnamefont\endcsname\relax
  \def\bibfnamefont#1{#1}\fi
\expandafter\ifx\csname citenamefont\endcsname\relax
  \def\citenamefont#1{#1}\fi
\expandafter\ifx\csname url\endcsname\relax
  \def\url#1{\texttt{#1}}\fi
\expandafter\ifx\csname urlprefix\endcsname\relax\def\urlprefix{URL }\fi
\providecommand{\bibinfo}[2]{#2}
\providecommand{\eprint}[2][]{\url{#2}}

\bibitem[{\citenamefont{Luttinger}(1964)}]{Luttinger64}
\bibinfo{author}{\bibfnamefont{J.~M.} \bibnamefont{Luttinger}},
  \bibinfo{journal}{Phys. Rev.} \textbf{\bibinfo{volume}{135}},
  \bibinfo{pages}{A1505} (\bibinfo{year}{1964}).

\bibitem[{\citenamefont{Shastry}(2009)}]{Shastry09}
\bibinfo{author}{\bibfnamefont{B.~S.} \bibnamefont{Shastry}},
  \bibinfo{journal}{Rep. Prog. Phys.} \textbf{\bibinfo{volume}{72}},
  \bibinfo{pages}{016501} (\bibinfo{year}{2009}).

\bibitem[{\citenamefont{Michaeli and Finkel'stein}(2009)}]{Michaeli09}
\bibinfo{author}{\bibfnamefont{K.}~\bibnamefont{Michaeli}} \bibnamefont{and}
  \bibinfo{author}{\bibfnamefont{A.~M.} \bibnamefont{Finkel'stein}},
  \bibinfo{journal}{Phys. Rev. B} \textbf{\bibinfo{volume}{80}},
  \bibinfo{pages}{115111} (\bibinfo{year}{2009}).

\bibitem[{\citenamefont{Schwiete and
  Finkel'stein}(2014{\natexlab{a}})}]{Schwiete14a}
\bibinfo{author}{\bibfnamefont{G.}~\bibnamefont{Schwiete}} \bibnamefont{and}
  \bibinfo{author}{\bibfnamefont{A.~M.} \bibnamefont{Finkel'stein}},
  \bibinfo{journal}{Phys. Rev. B} \textbf{\bibinfo{volume}{90}},
  \bibinfo{pages}{060201} (\bibinfo{year}{2014}{\natexlab{a}}).

\bibitem[{\citenamefont{Schwiete and
  Finkel'stein}(2014{\natexlab{b}})}]{Schwiete14b}
\bibinfo{author}{\bibfnamefont{G.}~\bibnamefont{Schwiete}} \bibnamefont{and}
  \bibinfo{author}{\bibfnamefont{A.~M.} \bibnamefont{Finkel'stein}},
  \bibinfo{journal}{Phys. Rev. B} \textbf{\bibinfo{volume}{90}},
  \bibinfo{pages}{155441} (\bibinfo{year}{2014}{\natexlab{b}}).

\bibitem[{\citenamefont{Castellani et~al.}(1987)\citenamefont{Castellani,
  Di~Castro, Kotliar, Lee, and Strinati}}]{Castellani87}
\bibinfo{author}{\bibfnamefont{C.}~\bibnamefont{Castellani}},
  \bibinfo{author}{\bibfnamefont{C.}~\bibnamefont{Di~Castro}},
  \bibinfo{author}{\bibfnamefont{G.}~\bibnamefont{Kotliar}},
  \bibinfo{author}{\bibfnamefont{P.~A.} \bibnamefont{Lee}}, \bibnamefont{and}
  \bibinfo{author}{\bibfnamefont{G.}~\bibnamefont{Strinati}},
  \bibinfo{journal}{Phys. Rev. Lett.} \textbf{\bibinfo{volume}{59}},
  \bibinfo{pages}{477} (\bibinfo{year}{1987}).

\bibitem[{\citenamefont{Tanatar et~al.}(2007)\citenamefont{Tanatar, Paglione,
  Petrovic, and Taillefer}}]{Tanatar07}
\bibinfo{author}{\bibfnamefont{M.~A.} \bibnamefont{Tanatar}},
  \bibinfo{author}{\bibfnamefont{J.}~\bibnamefont{Paglione}},
  \bibinfo{author}{\bibfnamefont{C.}~\bibnamefont{Petrovic}}, \bibnamefont{and}
  \bibinfo{author}{\bibfnamefont{L.}~\bibnamefont{Taillefer}},
  \bibinfo{journal}{Science} \textbf{\bibinfo{volume}{316}},
  \bibinfo{pages}{1320} (\bibinfo{year}{2007}).

\bibitem[{\citenamefont{Smith et~al.}(2008)\citenamefont{Smith, Sutherland,
  Lonzarich, Saxena, Kimura, Takashima, Nohara, and Takagi}}]{Smith08}
\bibinfo{author}{\bibfnamefont{R.~P.} \bibnamefont{Smith}},
  \bibinfo{author}{\bibfnamefont{M.}~\bibnamefont{Sutherland}},
  \bibinfo{author}{\bibfnamefont{G.~G.} \bibnamefont{Lonzarich}},
  \bibinfo{author}{\bibfnamefont{S.~S.} \bibnamefont{Saxena}},
  \bibinfo{author}{\bibfnamefont{N.}~\bibnamefont{Kimura}},
  \bibinfo{author}{\bibfnamefont{S.}~\bibnamefont{Takashima}},
  \bibinfo{author}{\bibfnamefont{M.}~\bibnamefont{Nohara}}, \bibnamefont{and}
  \bibinfo{author}{\bibfnamefont{H.}~\bibnamefont{Takagi}},
  \bibinfo{journal}{Nature} \textbf{\bibinfo{volume}{455}},
  \bibinfo{pages}{1220} (\bibinfo{year}{2008}).

\bibitem[{\citenamefont{Pfau et~al.}(2012)\citenamefont{Pfau, Hartmann,
  Stockert, Sun, Lausberg, Brando, Friedemann, Krellner, Geibel, Wirth
  et~al.}}]{Pfau12}
\bibinfo{author}{\bibfnamefont{H.}~\bibnamefont{Pfau}},
  \bibinfo{author}{\bibfnamefont{S.}~\bibnamefont{Hartmann}},
  \bibinfo{author}{\bibfnamefont{U.}~\bibnamefont{Stockert}},
  \bibinfo{author}{\bibfnamefont{P.}~\bibnamefont{Sun}},
  \bibinfo{author}{\bibfnamefont{S.}~\bibnamefont{Lausberg}},
  \bibinfo{author}{\bibfnamefont{M.}~\bibnamefont{Brando}},
  \bibinfo{author}{\bibfnamefont{S.}~\bibnamefont{Friedemann}},
  \bibinfo{author}{\bibfnamefont{C.}~\bibnamefont{Krellner}},
  \bibinfo{author}{\bibfnamefont{C.}~\bibnamefont{Geibel}},
  \bibinfo{author}{\bibfnamefont{S.}~\bibnamefont{Wirth}},
  \bibnamefont{et~al.}, \bibinfo{journal}{Nature}
  \textbf{\bibinfo{volume}{484}}, \bibinfo{pages}{493} (\bibinfo{year}{2012}).

\bibitem[{\citenamefont{Pfau et~al.}(2013)\citenamefont{Pfau, Daou, Lausberg,
  Naren, Brando, Friedemann, Wirth, Westerkamp, Stockert, Gegenwart
  et~al.}}]{Pfau13}
\bibinfo{author}{\bibfnamefont{H.}~\bibnamefont{Pfau}},
  \bibinfo{author}{\bibfnamefont{R.}~\bibnamefont{Daou}},
  \bibinfo{author}{\bibfnamefont{S.}~\bibnamefont{Lausberg}},
  \bibinfo{author}{\bibfnamefont{H.~R.} \bibnamefont{Naren}},
  \bibinfo{author}{\bibfnamefont{M.}~\bibnamefont{Brando}},
  \bibinfo{author}{\bibfnamefont{S.}~\bibnamefont{Friedemann}},
  \bibinfo{author}{\bibfnamefont{S.}~\bibnamefont{Wirth}},
  \bibinfo{author}{\bibfnamefont{T.}~\bibnamefont{Westerkamp}},
  \bibinfo{author}{\bibfnamefont{U.}~\bibnamefont{Stockert}},
  \bibinfo{author}{\bibfnamefont{P.}~\bibnamefont{Gegenwart}},
  \bibnamefont{et~al.}, \bibinfo{journal}{Phys. Rev. Lett.}
  \textbf{\bibinfo{volume}{110}}, \bibinfo{pages}{256403}
  (\bibinfo{year}{2013}).

\bibitem[{\citenamefont{Mahajan et~al.}(2013)\citenamefont{Mahajan, Barkeshli,
  and Hartnoll}}]{Mahajan13}
\bibinfo{author}{\bibfnamefont{R.}~\bibnamefont{Mahajan}},
  \bibinfo{author}{\bibfnamefont{M.}~\bibnamefont{Barkeshli}},
  \bibnamefont{and} \bibinfo{author}{\bibfnamefont{S.~A.}
  \bibnamefont{Hartnoll}}, \bibinfo{journal}{Phys. Rev. B}
  \textbf{\bibinfo{volume}{88}}, \bibinfo{pages}{125107}
  (\bibinfo{year}{2013}).

\bibitem[{\citenamefont{Dong et~al.}(2013)\citenamefont{Dong, Tokiwa, Bud'ko,
  Canfield, and Gegenwart}}]{Dong13}
\bibinfo{author}{\bibfnamefont{J.~K.} \bibnamefont{Dong}},
  \bibinfo{author}{\bibfnamefont{Y.}~\bibnamefont{Tokiwa}},
  \bibinfo{author}{\bibfnamefont{S.~L.} \bibnamefont{Bud'ko}},
  \bibinfo{author}{\bibfnamefont{P.~C.} \bibnamefont{Canfield}},
  \bibnamefont{and}
  \bibinfo{author}{\bibfnamefont{P.}~\bibnamefont{Gegenwart}},
  \bibinfo{journal}{Phys. Rev. Lett.} \textbf{\bibinfo{volume}{110}},
  \bibinfo{pages}{176402} (\bibinfo{year}{2013}).

\bibitem[{\citenamefont{Sutherland et~al.}(2015)\citenamefont{Sutherland,
  O'Farrell, Toews, Dunn, Kuga, Nakatsuji, Machida, Izawa, and
  Hill}}]{Sutherland15}
\bibinfo{author}{\bibfnamefont{M.~L.} \bibnamefont{Sutherland}},
  \bibinfo{author}{\bibfnamefont{E.~C.~T.} \bibnamefont{O'Farrell}},
  \bibinfo{author}{\bibfnamefont{W.~H.} \bibnamefont{Toews}},
  \bibinfo{author}{\bibfnamefont{J.}~\bibnamefont{Dunn}},
  \bibinfo{author}{\bibfnamefont{K.}~\bibnamefont{Kuga}},
  \bibinfo{author}{\bibfnamefont{S.}~\bibnamefont{Nakatsuji}},
  \bibinfo{author}{\bibfnamefont{Y.}~\bibnamefont{Machida}},
  \bibinfo{author}{\bibfnamefont{K.}~\bibnamefont{Izawa}}, \bibnamefont{and}
  \bibinfo{author}{\bibfnamefont{R.~W.} \bibnamefont{Hill}},
  \bibinfo{journal}{Phys. Rev. B} \textbf{\bibinfo{volume}{92}},
  \bibinfo{pages}{041114} (\bibinfo{year}{2015}).

\bibitem[{\citenamefont{Taupin et~al.}(2015)\citenamefont{Taupin, Knebel,
  Matsuda, Lapertot, Machida, Izawa, Brison, and Flouquet}}]{Taupin15}
\bibinfo{author}{\bibfnamefont{M.}~\bibnamefont{Taupin}},
  \bibinfo{author}{\bibfnamefont{G.}~\bibnamefont{Knebel}},
  \bibinfo{author}{\bibfnamefont{T.~D.} \bibnamefont{Matsuda}},
  \bibinfo{author}{\bibfnamefont{G.}~\bibnamefont{Lapertot}},
  \bibinfo{author}{\bibfnamefont{Y.}~\bibnamefont{Machida}},
  \bibinfo{author}{\bibfnamefont{K.}~\bibnamefont{Izawa}},
  \bibinfo{author}{\bibfnamefont{J.-P.} \bibnamefont{Brison}},
  \bibnamefont{and} \bibinfo{author}{\bibfnamefont{J.}~\bibnamefont{Flouquet}},
  \bibinfo{journal}{Phys. Rev. Lett.} \textbf{\bibinfo{volume}{115}},
  \bibinfo{pages}{046402} (\bibinfo{year}{2015}).

\bibitem[{\citenamefont{Wiedemann and Franz}(1853)}]{Wiedemann1853}
\bibinfo{author}{\bibfnamefont{G.}~\bibnamefont{Wiedemann}} \bibnamefont{and}
  \bibinfo{author}{\bibfnamefont{R.}~\bibnamefont{Franz}},
  \bibinfo{journal}{Ann.~Phys. (Leipzig)} \textbf{\bibinfo{volume}{89}},
  \bibinfo{pages}{497} (\bibinfo{year}{1853}).

\bibitem[{\citenamefont{Livanov et~al.}(1991)\citenamefont{Livanov, Reizer, and
  Sergeev}}]{Livanov91}
\bibinfo{author}{\bibfnamefont{D.~V.} \bibnamefont{Livanov}},
  \bibinfo{author}{\bibfnamefont{M.}~\bibnamefont{Reizer}}, \bibnamefont{and}
  \bibinfo{author}{\bibfnamefont{A.~V.} \bibnamefont{Sergeev}},
  \bibinfo{journal}{Zh.~Eksp.~Teor.~Fiz.} \textbf{\bibinfo{volume}{99}},
  \bibinfo{pages}{1230} (\bibinfo{year}{1991}), \bibinfo{note}{[Sov.~Phys.~JETP
  {\bf 72}, 760 (1991)]}.

\bibitem[{\citenamefont{Arfi}(1992)}]{Arfi92}
\bibinfo{author}{\bibfnamefont{B.}~\bibnamefont{Arfi}}, \bibinfo{journal}{J.
  Low Temp. Phys.} \textbf{\bibinfo{volume}{86}}, \bibinfo{pages}{213}
  (\bibinfo{year}{1992}).

\bibitem[{\citenamefont{Raimondi et~al.}(2004)\citenamefont{Raimondi, Savona,
  Schwab, and L\"uck}}]{Raimondi04}
\bibinfo{author}{\bibfnamefont{R.}~\bibnamefont{Raimondi}},
  \bibinfo{author}{\bibfnamefont{G.}~\bibnamefont{Savona}},
  \bibinfo{author}{\bibfnamefont{P.}~\bibnamefont{Schwab}}, \bibnamefont{and}
  \bibinfo{author}{\bibfnamefont{T.}~\bibnamefont{L\"uck}},
  \bibinfo{journal}{Phys. Rev. B} \textbf{\bibinfo{volume}{70}},
  \bibinfo{pages}{155109} (\bibinfo{year}{2004}).

\bibitem[{\citenamefont{Niven and Smith}(2005)}]{Niven05}
\bibinfo{author}{\bibfnamefont{D.~R.} \bibnamefont{Niven}} \bibnamefont{and}
  \bibinfo{author}{\bibfnamefont{R.~A.} \bibnamefont{Smith}},
  \bibinfo{journal}{Phys. Rev. B} \textbf{\bibinfo{volume}{71}},
  \bibinfo{pages}{035106} (\bibinfo{year}{2005}).

\bibitem[{\citenamefont{Catelani and Aleiner}(2005)}]{Catelani05}
\bibinfo{author}{\bibfnamefont{G.}~\bibnamefont{Catelani}} \bibnamefont{and}
  \bibinfo{author}{\bibfnamefont{I.~L.} \bibnamefont{Aleiner}},
  \bibinfo{journal}{Zh. Eksp. Teor. Fiz.} \textbf{\bibinfo{volume}{127}},
  \bibinfo{pages}{327} (\bibinfo{year}{2005}), \bibinfo{note}{[Sov.~Phys.~JETP
  {\bf 100}, 331 (2005)]}.

\bibitem[{\citenamefont{Catelani}(2007)}]{Catelani07}
\bibinfo{author}{\bibfnamefont{G.}~\bibnamefont{Catelani}},
  \bibinfo{journal}{Phys. Rev. B} \textbf{\bibinfo{volume}{75}},
  \bibinfo{pages}{024208} (\bibinfo{year}{2007}).

\bibitem[{\citenamefont{Belinfante}(1940)}]{Belinfante40}
\bibinfo{author}{\bibfnamefont{F.~J.} \bibnamefont{Belinfante}},
  \bibinfo{journal}{Physica} \textbf{\bibinfo{volume}{7}}, \bibinfo{pages}{449
  } (\bibinfo{year}{1940}).

\bibitem[{\citenamefont{Greiner and Reinhardt}(1996)}]{Greiner96}
\bibinfo{author}{\bibfnamefont{W.}~\bibnamefont{Greiner}} \bibnamefont{and}
  \bibinfo{author}{\bibfnamefont{J.}~\bibnamefont{Reinhardt}},
  \emph{\bibinfo{title}{Field Quantization}}
  (\bibinfo{publisher}{Springer-Verlag}, \bibinfo{address}{Berlin},
  \bibinfo{year}{1996}).

\bibitem[{\citenamefont{Schwinger}(1961)}]{Schwinger61}
\bibinfo{author}{\bibfnamefont{J.}~\bibnamefont{Schwinger}},
  \bibinfo{journal}{J.~Math.~Phys.} \textbf{\bibinfo{volume}{2}},
  \bibinfo{pages}{407} (\bibinfo{year}{1961}).

\bibitem[{\citenamefont{Kadanoff and Baym}(1962)}]{Kadanoff62}
\bibinfo{author}{\bibfnamefont{L.~P.} \bibnamefont{Kadanoff}} \bibnamefont{and}
  \bibinfo{author}{\bibfnamefont{G.}~\bibnamefont{Baym}},
  \emph{\bibinfo{title}{Quantum Statistical Mechanics}} (\bibinfo{publisher}{W.
  A. Benjamin, New York}, \bibinfo{year}{1962}).

\bibitem[{\citenamefont{Keldysh}(1964)}]{Keldish65}
\bibinfo{author}{\bibfnamefont{L.~V.} \bibnamefont{Keldysh}},
  \bibinfo{journal}{Zh.~Eksp.~Teor.~Fiz.} \textbf{\bibinfo{volume}{47}},
  \bibinfo{pages}{1515} (\bibinfo{year}{1964}), \bibinfo{note}{[Sov.~Phys.~JETP
  {\bf 20}, 1018 (1965)]}.

\bibitem[{\citenamefont{Kamenev}(2011)}]{Kamenev11}
\bibinfo{author}{\bibfnamefont{A.}~\bibnamefont{Kamenev}},
  \emph{\bibinfo{title}{Non-Equilibrium Systems}}
  (\bibinfo{publisher}{Cambridge University Press, Cambridge},
  \bibinfo{year}{2011}).

\bibitem[{\citenamefont{Larkin and Ovchinnikov}(1975)}]{Larkin75}
\bibinfo{author}{\bibfnamefont{A.~I.} \bibnamefont{Larkin}} \bibnamefont{and}
  \bibinfo{author}{\bibfnamefont{Y.~N.} \bibnamefont{Ovchinnikov}},
  \bibinfo{journal}{Zh.~Eksp.~Teor.~Fiz.} \textbf{\bibinfo{volume}{68}},
  \bibinfo{pages}{1915} (\bibinfo{year}{1975}), \bibinfo{note}{[Sov.~Phys.~JETP
  {\bf 41}, 960 (1975)]}.

\bibitem[{\citenamefont{Schwiete and
  Finkel'stein}(2014{\natexlab{c}})}]{Schwiete14}
\bibinfo{author}{\bibfnamefont{G.}~\bibnamefont{Schwiete}} \bibnamefont{and}
  \bibinfo{author}{\bibfnamefont{A.~M.} \bibnamefont{Finkel'stein}},
  \bibinfo{journal}{Phys. Rev. B} \textbf{\bibinfo{volume}{89}},
  \bibinfo{pages}{075437} (\bibinfo{year}{2014}{\natexlab{c}}).

\bibitem[{\citenamefont{Finkel'stein}(1983)}]{Finkelstein83}
\bibinfo{author}{\bibfnamefont{A.~M.} \bibnamefont{Finkel'stein}},
  \bibinfo{journal}{Zh. Exp. Teor. Fiz.} \textbf{\bibinfo{volume}{84}},
  \bibinfo{pages}{168} (\bibinfo{year}{1983}), \bibinfo{note}{[Sov. Phys. JETP
  {\bf 57}, 97 (1983)]}.

\bibitem[{\citenamefont{Castellani et~al.}(1984)\citenamefont{Castellani,
  Di~Castro, Lee, and Ma}}]{Castellani84}
\bibinfo{author}{\bibfnamefont{C.}~\bibnamefont{Castellani}},
  \bibinfo{author}{\bibfnamefont{C.}~\bibnamefont{Di~Castro}},
  \bibinfo{author}{\bibfnamefont{P.~A.} \bibnamefont{Lee}}, \bibnamefont{and}
  \bibinfo{author}{\bibfnamefont{M.}~\bibnamefont{Ma}}, \bibinfo{journal}{Phys.
  Rev. B} \textbf{\bibinfo{volume}{30}}, \bibinfo{pages}{527}
  (\bibinfo{year}{1984}).

\bibitem[{\citenamefont{Di~Castro and Raimondi}(2004)}]{DiCastro04}
\bibinfo{author}{\bibfnamefont{C.}~\bibnamefont{Di~Castro}} \bibnamefont{and}
  \bibinfo{author}{\bibfnamefont{R.}~\bibnamefont{Raimondi}}, in
  \emph{\bibinfo{booktitle}{The electron liquid paradigm in condensed matter
  physics: Proceedings of the International School of Physics "Enrico Fermi":
  Varenna, Italy, 29 July-8 August 2003}}, edited by
  \bibinfo{editor}{\bibfnamefont{G.~F.} \bibnamefont{Giuliani}}
  \bibnamefont{and} \bibinfo{editor}{\bibfnamefont{G.}~\bibnamefont{Vignale}}
  (\bibinfo{publisher}{IOS Press, Amsterdam}, \bibinfo{year}{2004}), pp.
  \bibinfo{pages}{259--333}.

\bibitem[{\citenamefont{Finkel'stein}(2010)}]{Finkelstein10}
\bibinfo{author}{\bibfnamefont{A.~M.} \bibnamefont{Finkel'stein}}, in
  \emph{\bibinfo{booktitle}{50 years of Anderson Localization}}, edited by
  \bibinfo{editor}{\bibfnamefont{E.}~\bibnamefont{Abrahams}}
  (\bibinfo{publisher}{World Scientific Publishing Co.},
  \bibinfo{address}{Singapore}, \bibinfo{year}{2010}), p. \bibinfo{pages}{385}.

\bibitem[{\citenamefont{Castellani and Di~Castro}(1986)}]{Castellani86}
\bibinfo{author}{\bibfnamefont{C.}~\bibnamefont{Castellani}} \bibnamefont{and}
  \bibinfo{author}{\bibfnamefont{C.}~\bibnamefont{Di~Castro}},
  \bibinfo{journal}{Phys. Rev. B} \textbf{\bibinfo{volume}{34}},
  \bibinfo{pages}{5935} (\bibinfo{year}{1986}).

\bibitem[{\citenamefont{Lyakhov and Mishchenko}(2003)}]{Lyakhov03}
\bibinfo{author}{\bibfnamefont{A.~O.} \bibnamefont{Lyakhov}} \bibnamefont{and}
  \bibinfo{author}{\bibfnamefont{E.~G.} \bibnamefont{Mishchenko}},
  \bibinfo{journal}{Phys. Rev. B} \textbf{\bibinfo{volume}{67}},
  \bibinfo{pages}{041304} (\bibinfo{year}{2003}).

\bibitem[{\citenamefont{Jackson}(1975)}]{Jackson75}
\bibinfo{author}{\bibfnamefont{J.~D.} \bibnamefont{Jackson}},
  \emph{\bibinfo{title}{Classical Electrodynamics}} (\bibinfo{publisher}{Wiley,
  New York}, \bibinfo{year}{1975}).

\bibitem[{\citenamefont{Altshuler and Aronov}(1985)}]{Altshuler85}
\bibinfo{author}{\bibfnamefont{B.~L.} \bibnamefont{Altshuler}}
  \bibnamefont{and} \bibinfo{author}{\bibfnamefont{A.~G.}
  \bibnamefont{Aronov}}, \emph{\bibinfo{title}{Electron--Electron Interaction
  in Disordered Conductors}} (\bibinfo{publisher}{North--Holland},
  \bibinfo{address}{Amsterdam}, \bibinfo{year}{1985}),
  vol.~\bibinfo{volume}{10} of \emph{\bibinfo{series}{Modern Problems in
  Condensed Matter Sciences}}, pp. \bibinfo{pages}{1--153}.

\end{thebibliography}

\end{document}